\newcounter{secnum}

\newcommand{\mysection}[1]{%
\vspace{1.25\baselineskip}
\stepcounter{section}
\stepcounter{secnum}
\centerline{\large\bf\thesecnum. #1}
\vspace{1pt}
}

\newcounter{subsecnum}[secnum]

\newcommand{\mysubsection}[1]{%
\vspace{0.5\baselineskip}
\stepcounter{subsecnum}
\hbox{\bf\Alph{subsecnum}. #1}
\vspace{1pt}
}

\newenvironment{morelines}{\bgroup\let\\=\cr\vbox\bgroup
\ialign\bgroup\hfil$\displaystyle##$&
\hfil$\displaystyle\null##\null$\hfil&
$\displaystyle##\hfil$\cr}{\cr\egroup\egroup\egroup}

\newcommand{\fig}[1]{Fig.~#1}

\newcommand{\psibar}[1]{#1\psi\bar{\psi}}
\newcommand{\psiV}{\psibar{V}}

\newcommand{\half}{\frac{1}{2}}

\newbox\slashbox
\newdimen\slashwd
\newcommand{\slashed}[1]{%
\setbox\slashbox=\hbox{$#1$}%
\slashwd=\wd\slashbox%
\hbox to\slashwd{\hss/\hss}%
\llap{$#1$}}

\newcommand{\Deltaslash}{\slashed{\Delta}}
\newcommand{\kslash}{\slashed{k}}
\newcommand{\pslash}{\slashed{p}}
\newcommand{\Pslash}{\slashed{P}}
\newcommand{\qslash}{\slashed{q}}

\documentstyle[12pt]{article}

\begin{document}
 
\begin{titlepage}
\begin{center}
{\bf HOW TO TREAT $\gamma_{5}$}\\
\end{center}

\begin{center}
Hung Cheng$^1$\\
Department of Mathematics, Massachusetts Institute of Technology\\
Cambridge, MA  02139, U.S.A.\\

\bigskip
and
\bigskip

S.P. Li$^2$\\
Institute of Physics, Academia Sinica\\
Nankang, Taipei, Taiwan, Republic of China\\
\end{center}



\noindent
1. E-mail: huncheng@math.mit.edu

\noindent
2. E-mail: spli@phys.sinica.edu.tw
 
\end{titlepage}

\begin{abstract}

  In this paper, we present a method to perform renormalized
  perturbation calculations in gauge theories with chiral fermions.
  We find it proper to focus directly on the Ward-Takahashi
  identities, relegating dimensional regularization into a
  supplementary and secondary role.  We show with the example of the
  Abelian-Higgs theory how to handle amplitudes involving fermions,
  particularly how to handle the matrix $\gamma_{5}$.  As a
  demonstration of our method of renormalization, we evaluate the
  radiative corrections of the triangular anomaly in the Abelian-Higgs
  theory with chiral fermions and with Yukawa couplings. Families of
  chiral fermions with appropriate quantum numbers are introduced so
  that the sum of their contributions to the lowest-order anomaly is
  equal to zero.  The left-handed and right-handed fermion of this
  theory are assumed to interact with the gauge field in the mixture
  $\displaystyle (1+\theta)\frac{(1+\gamma_5)}{2} + \theta \frac {(1-
\gamma_5)}{2} $ where $\theta$ is a constant.  This anomaly amplitude
  is calculated without any regularization and is found to vanish. 

\end{abstract}

\newpage

\mysection{Introduction}

Ever since the early seventies, it has become customary to renormalize
quantum gauge field theories with dimensional regularization, with the 
classic method of renormalization on the basis of subtractions with the
aid of the Ward-Takahashi identities largely ignored.  In this paper,
we shall re-examine some basic issues in the theory of renormalization,
particularly as applied to quantum gauge field theories with
spontaneously broken symmetries and with chiral fermions.
As we all know, a distinctive feature of a quantum gauge field theory
is that the number of renormalized parameters often exceeds that
of the bare parameters. As a general rule, a quantum field theory with
an excessive number of renormalized parameters is likely to be not
renormalizable. Take, for example, the theory of scalar QED.  It is
well-known that, if we follow spinor QED to the letter and introduce
only two bare parameters, the bare charge and the bare mass of the
scalar meson, the corresponding quantum field theory is not
renormalizable. This is because there are three parameters which must
be renormalized, the third one being the $|\phi^4|$ coupling
constant. In order to make scalar electrodynamics renormalizable, one
additional parameter, the unrenormalized $|\phi^4|$ coupling constant,
must be introduced. It is notable that quantum gauge field theories
with a spontaneous broken vacuum symmetry are exceptions to the general
rule. Take the example of the Abelian-Higgs field theory.  There are in
the theory three bare constants in the boson sector: the bare
coupling of the gauge meson, and the two bare parameters in the Higgs
potential---same as in scalar QED. These three bare constants generated 
more than three
renormalized parameters: the physical masses of the gauge meson and
the Higgs meson, the renormalized coupling constants of the gauge
meson, and the renormalized 3-point and 4-point coupling constants of
the scalar meson.  Of particular importance is the physical mass of
the gauge meson.  Trivially, the mass of the photon in scalar QED
is finite, as it is equal to zero.  In contrast, the mass of the gauge
meson in the Abelion-Higgs field theory is non-zero.  In order to
prove that the Abelian-Higgs theory is renormalizable, one must show
that this extra renormalized parameter of mass is finite to all orders 
of perturbation.  This can be done by showing that the gauge meson
is related to the other renormalized parameters.  Such an explicit
non-perturbative relation$^{1}$, as well as
others which put contraints on the renormalized parameters, are
provided by the Ward-Takahashi identities.

The contents of the Ward-Takahashi identities in a gauge field theory with
spontaneous symmetry breaking are complicated and require some care to
disentangle. In order to sidestep these complications, the method of
dimensional regularization has been invented $^{2,3}$. With the use of
this method, the renormalized perturbation series automatically
satisfies the Ward-Takahashi identities, and one may be led to believe that there
is no more need to pay a great deal of attention to the identities which
have already been incorporated.

We consider such a belief misplaced. While the method of dimensional
regularization is convenient to use in explicit calculations, there are
two shortcomings associated with it. First, the definition of the matrix
$\gamma_{5}$ for a non-integral dimension is subjective and controversial up to now\footnotemark[4].
Therefore, the renormalization theory on the basis of dimensional
regularization alone remains incomplete.  Second, and perhaps more
important, the Ward-Takahashi identities contain rich and physical implications
which should be explored. Thus, ignoring the Ward-Takahashi identities under the
auspices of dimensional regularization is not an unmitigated blessing.

We believe that the Ward-Takahashi identities are the centerpieces of
renormalization. Their contents should be extracted and utilized. The
method of dimensional regularization, on the other hand, should be
recognized as it is: a mathematical artifice which is helpful to use in
some cases---no more and no less.

\mysection{Subtractions in the Fermion Sector}

While the renormalization procedure for the fermion propagator and the
fermion vertex functions in QED is well-known, new features appear
when chiral fermions enter the picture. Unlike QED, a theory for
chiral fermions has no parity conservation.  By Lorentz covariance,
this 1PI amplitude is of the form $$
\delta m + ma(p^2) +m \gamma_5b(p^2)+c(p^2)\pslash+d(p^2)\pslash\gamma_5
\eqno{(2.1)}
$$
where 
$$
\delta m \equiv m-m_0.
$$
The unrenormalized propagator $S(p)$ is therefore given by
$$
S(p) = i\,\bigl[ \pslash(1-c)-m(1+a)-(bm+d\pslash)\gamma_5 \bigr]^{-1},
\eqno{(2.2)}
$$
which can be shown to be equal to
$$
S(p) =i\,
\frac{\pslash(1-c)+m(1+a)-(bm+d\pslash)\gamma_5}{p^2(1-c-d)(1-c+d)-m^2(1+a+b)(1+a-b)}.
\eqno{(2.3)}
$$
 
The invariant amplitudes $b$ and $d$ are absent in QED, hence there are,
in a quantum field theory with chiral fermions, two more invariant
functions which must be rendered finite by renormalization.  That all 
these functions are finite is due to two additional subtraction 
conditions provided by the hypothesis of the existence of in-fields and
out-fields.  To see this, we note that
the matrix elements of $S(p)$ are given by
$$
S_{\alpha\beta}(p) = \int d^4(x-y)e^{ip(x-y)} <
0|T\psi_{\alpha}(x)\bar{\psi}_{\beta}(y)|0>.
\eqno{(2.4)}
$$
By assumption, it has a pole at
$$
p^2_0 = E^2,
$$
where
$$
E \equiv \sqrt{\vec{p}^2+m^2.}
$$
Let
$$
R_{\alpha\beta} \equiv \lim_{x_0\rightarrow\infty \atop y_0\rightarrow-\infty}
\int d^3(\vec{x}-\vec{y})e^{-i\vec{p}.(\vec{x}-\vec{y})} <
0|\psi_\alpha(x)\bar{\psi}_\beta(y)|0>.
\eqno{(2.5)}
$$
As $y_0 \rightarrow -\infty, \bar{\psi}_\beta(y)$ turns into the in-field and
we get
$$
\bar{\psi}_\beta(y)|0>\ \rightarrow \int \frac{d^3p}{(2\pi)^3}e^{ip.y}
\sqrt{\frac{m}{E(\vec{p})}} a^+_{in}(\vec{p})\bar{u}_\beta(\vec{p})|0>
$$
where the summation over the two spinor states are not exhibited.  Also,
$a^+_{in}(\vec{p})$ is the creation operator for an incoming dressed fermion,
and $u(\vec{p})$ is the spinor wavefunction for a fermion satisfying
$$
\pslash u(\vec{p}) = mu(\vec{p}).
$$
Similarly, as $x_0 \rightarrow \infty, \psi_\alpha(x)$ turns into the out-field
and we get
$$
<0|\psi_\alpha(x) \rightarrow \int \frac{d^3p}{(2\pi)^3}e^{-ip.x}
\sqrt{\frac{m}{E(\vec{p})}} < 0|u_\alpha(\vec{p})a_{out}(\vec{p}).
$$
The matrix $R$ therefore satisfies
$$
\bar{v}(\vec{p})Rv(\vec{p}) = \bar{v}(\vec{p})Ru(\vec{p}) =
\bar{u}(\vec{p})Rv(\vec{p})=0,
\eqno{(2.6)}
$$
where $v(\vec{p})$ is a spinor wavefunction for an antifermion satisfying
$$
\pslash v(\vec{p}) = -mv(\vec{p}).
$$
Equation (2.6) is obtained as
$$
\bar{v}(\vec{p})u(\vec{p})=\bar{u}(\vec{p})v(\vec{p})=0.
$$

The pole of $S(p)$ at $p_0=E$ comes from the integration over very large
positive values of $(x_0-y_0)$, as the finite range of $(x_0-y_0)$ cannot
contribute a value of infinity.  Since
$$
\int^\infty_0 dte^{i(p_0-E)t} = \frac{i}{p_0-E}, 
\eqno{(2.7)}
$$
iR is the residue of $S(p)$ at $p_0=E$.  Requiring that the numerator in (2.3) 
at $p_0 = E$ satisfies (2.6),
we get
$$
c(m^2) = -a(m^2),
\eqno{(2.8)}
$$
$$
d(m^2) = 0,
\eqno{(2.9a)}
$$
and
$$
b(m^2)=0,
\eqno{(2.9b)}
$$
which are the subtraction conditions needed for the divergent amplitudes
$a,b,c$ and $d$.

From (2.3), we find that, when $p^2 \approx m^2$ and $\pslash$ is set to m,
$$
S(p) \approx \frac{2mi}{p^2-m^2}Z_{\psi}(m)
\eqno{(2.10)}
$$
where 
$$
Z_\psi(m) \equiv \frac{1}{1-c(m^2)-2m^2[c'(m^2)+a'(m^2)]}.
\eqno{(2.11)}
$$
Eq.(2.11) is obtained by differentiating the denominator of the right-hand
side of (2.3) and utilizing (2.8) and (2.9).  

Let us define
$$
S^{(r)}(p) \equiv S(p)/Z_\psi(m),
\eqno{(2.12)}
$$
then the renormalized fermion propagator $S^{(r)}(p)$ is
approximately equal to
$$
\frac{2mi}{p^2-m^2}
\eqno{(2.13)}
$$
when $p^2$ is approximately equal to $m^2$ and $\pslash$ is set to m.

In a perturbative calculation, the 1PI amplitude for the fermion
propagator is linearly divergent.  As we know, a shift of the momentum
variable of a linearly divergent integral gives birth to a finite term.
Therefore, we may interpret a Feynman integral of the 1PI amplitude of
the fermion propagator as one of symmetric integration, with an unknown
additive constant arising from an undetermined amount of shift of the
momentum variables.  The symmetrically integrated amplitude is
logarithmically divergent.

There are two observations: (i) Because the amplitudes $a$ and $b$ in
(2.1) are multiplied by a factor of $m$, these
amplitudes are not linearly divergent by power counting.  
Therefore, they do not contain
unknown additive constants, which appear only in the amplitudes $c$ and
$d$, (ii) There are no counter terms in the Lagrangian for the
amplitudes $d$ and $b$.  Since the amplitude $b$ does not contain an
unknown additive constant, it must be finite and must satisfy (2.9b) on
its own.  The amplitude $d$, on the other hand, is allowed an additive
constant contributed by linearly divergent integrals.  This constant is
determined by (2.9a).

The renormalized perturbation series for the renormalized fermion
self-energy 1PI amplitude can therefore be obtained as follows.  For a
graph which has no divergent subgraphs, we employ the Feynman rules to
obtain perturbatively the amplitudes $a,b,c,$ and $d$ with the coupling
constants and masses being the renomalized ones.  The divergent integrals are
symmetrically integrated after Feynman parameters are introduced.  The
amplitude $ma(p^2)+c(p^2)\pslash$ remains to be logarithmically
divergent, and is replaced by the subtracted amplitude
$$
 m \bigl[ a(p^2)-a(m^2) \bigr] + \bigl[ c(p^2)-c(m^2) \bigr] \pslash
 -2m^2 \bigl[ a'(m^2)+c'(m^2) \bigr] (\pslash-m).
\eqno{(2.14)}
$$
The last term of (2.14) is obtained by requiring that the residue
of $S^{(r)}$ at $p^2 = m^2$ with $\pslash = m$ is $2mi$.
These subtractions are the same as the ones in QED.  
The amplitude $b(p^2)$
requires no subtraction, while the amplitude $d(p^2)$ is replaced by the
subtracted amplitude
$$
d(p^2)-d(m^2),
$$
so that (2.9a) is satisfied.  For graphs with divergent subgraphs, we use the
BPHZ formalism$^5$.  We shall express the renomalized 1PI amplitude for the
fermion propagator as
$$
m a_r(p^2)+m\gamma_5b_r(p^2)+c_r(p^2)\pslash+d_r(p^2)\pslash\gamma_5.
\eqno{(2.15a)}
$$
These renormalized functions are related to the unrenormalized functions by
$$\begin{array}{ll}
1+a_r & \equiv Z_\psi (m)(1+a), \\
1-c_r & \equiv Z_\psi (m)(1-c), \\ 
b_r & \equiv Z_\psi (m)b, \\
d_r & \equiv Z_\psi (m)d.
\end{array} 
\eqno{(2.15b)}
$$

We shall demonstrate our method of renormalization by the specific example 
of the Abelian Higgs model.
The Langrangian density in the Abelian-Higgs model is given by
$$\begin{array}{ll}
{\cal L} = & - \displaystyle \frac{1}{4} F_{\mu\nu} F^{\mu\nu} + (D_{\mu}\phi)^+ (D^\mu \phi) +
\sum_i \biggl[ \bar{\psi}_{Li}(i\partial\!\!\!/ - g_0 (1+\theta)V\!\!\!\!/) \psi_{Li}
 + \bar{\psi}_{Ri} (i\partial\!\!\!/ - g_0 \theta V\!\!\!\!/) \psi_{Ri} \\ 
& - \sqrt{2} f_{0i} \phi
\bar{\psi}_{Li} \psi_{Ri} - \sqrt{2} f_{0i} \phi^+ \bar{\psi}_{Ri} \psi_{Li} \biggr] + \mu^{2}_{0} \phi^{+}\phi - \lambda_0(\phi^{+}\phi)^2,
\end{array}
\eqno {(2.16)}
$$
with
$$
D_\mu \phi \equiv (\partial_\mu + ig_0 V_\mu)\phi, 
$$
$$
\psi_{Li} \equiv \frac{1}{2}(1+ \gamma_5)\psi_i, 
$$
$$
\psi_{Ri} \equiv \frac{1}{2}(1- \gamma_5)\psi_i, 
$$ and $$ F_{\mu\nu} \equiv \partial_\mu V_\nu - \partial_\nu V_\mu.
$$ In the above, $V_{\mu},\phi$, and $\psi_i$ are the gauge field, the
complex scalar field, and the ${i^{th}}$ chiral fermion field,
respectively, and $\theta, g_0, f_{0i}, \mu_0$ and $\lambda_0$ are
constants.  The subscript {0} of the constants in (2.16) signifies
that these constants are bare.  The summation in (2.16) is over all of
the chiral fermions we have introduced into the theory.  For
simplicity, we shall drop the subscript $i$ for the individual chiral
fermions from here onward but keep in mind that $f_0, \psi$, etc
refer to any one of the chiral fermions introduced.  As usual, we
shall put $$
\phi \equiv \frac{v_0 + H + i \phi_2}{\sqrt{2}}
$$
where
$
v_0 \equiv {\mu_0}/{\sqrt{\lambda_0}}.
$
We shall also call the bare mass of the gauge meson as
$
M_{0}\equiv g_0v_0,
$
and the bare mass of the fermion as 
$
m_0 \equiv f_0v_0.
$

We shall add to the Lagrangian (2.16) a gauge-fixing term and a ghost term.  Thus,
we consider the effective Lagrangian 
$$
L_{eff} \equiv {\cal L} - \frac{1}{2\alpha}\ell^2 -
i(\partial_{\mu}\eta)(\partial^{\mu}\xi)+i \alpha M^2_0 \eta\xi + i \alpha g_0
M_0 \eta\xi H, 
\eqno{(2.17)}
$$
where
$$
\ell = \partial_{\mu} V^{\mu} - \alpha M_0 \phi_2,
$$
$\alpha$ is a constant, and $\xi$ and $\eta$ are ghost fields.
The effective Lagrangian is invariant under the following BRST variations:
$$
\delta V_\mu = \partial_\mu \xi ,  \,\,\, \delta H = g_0 \xi \phi_2 , \,\,\,
\delta \phi_2 = - g_0 \xi (v_0 + H) ,
$$
$$
\displaystyle
\delta i \eta = \frac{1}{\alpha} \ell , \,\,\, \delta \xi = 0,
$$
$$
\delta \psi_L = - i g_0 ( 1+\theta) \xi \psi_L , \,\,\, \delta \psi_R = - i
g_0 \theta \xi \psi_R \, .
\eqno(2.18)
$$

Next we turn to the Ward-Takahashi identity for three-point functions involving
fermions.  Starting from$^1$
$$
<0|\delta Ti\eta(x)\psi(y)\bar{\psi}(z)|0>=0,
$$
where $\psi (x)$ is any of the chiral fermion fields.  We get
$$
<0|T(\frac{1}{\alpha}\partial_{\mu}V^{\mu}-M_0\phi_2)(\psi)(\bar{\psi})|0>\!\!-\!\!<0|(i\eta)
(-ig_0\xi\frac{1+2\theta+\gamma_5}{2}\psi)(\bar{\psi})|0>
$$
$$-\!\!<0|(i\eta)(\psi)(ig_0\xi\bar{\psi}\frac{1+2\theta-\gamma_5}{2})|0>\ =\ 0.
\eqno{(2.19)}
$$
We shall adopt the Landau gauge, taking the limit $\alpha\rightarrow 0$.  
Multiplying (2.15) by $\displaystyle e^{-i\Delta x+ip^\prime y-ip z}$ and
integrating, we get:
$$
-
i\Delta_\mu\Gamma^{\mu}_{V\psi\bar{\psi}}(p^\prime,p)
-
m_0 Z\Gamma_{\phi_2 \psi\bar{\psi}}(p^\prime,p)
=
S^{-1}(p^\prime)\frac{1+2\theta+\gamma_5}{2}
-
\frac{1+2\theta-\gamma_5}{2}S^{-1}(p)  ,
\eqno{(2.20)}$$
where $\displaystyle Z \equiv \frac{1}{\it{v}_0} < 0|(\it{v}_0 + H)|0> $.  
In (2.20), $p'$ and $p$ are the outgoing and incoming momentum
for the fermion, respectively, and 
$$
\Delta=p^\prime-p.
$$
The functions $\displaystyle\Gamma^{\mu}_{V\psi\bar{\psi}}$,
$\displaystyle\Gamma_{\phi_2\psi\bar{\psi}}$ and $S(p)$ are so defined that
their lowest-order terms in the unrenormalized perturbation
series are $\displaystyle\gamma^{\mu}\frac{1+2\theta+\gamma_5}{2}$,$\displaystyle-i\gamma_5$ and
$\displaystyle\frac{i}{\pslash-m}$, respectively.

If we set $\Delta=0$ in (2.20), we get
$$
m_0
Z\Gamma_{\phi_2\psi\bar{\psi}}(p,p)=-\frac{1}{2}[\gamma_5S^{-1}(p)+S^{-1}(p)\gamma_5] .
\eqno{(2.21)}
$$
Let us set, in addition,  $p^2=m^2$ in (2.21).  We get, referring to (2.2), 
$$\displaystyle
m_0 Z\Gamma_{\phi_2\psi\bar{\psi}}(p,p)|_{p^2=m^2}=-im[1+a(m^2)]\gamma_5 .
\eqno{(2.22)}
$$

By Lorentz covariance, $\displaystyle\Gamma_{\phi_2\psi\bar{\psi}}$ is a
superposition of scalar amplitudes and pseudo-scalar amplitudes:
$$\begin{array}{ll} 
\Gamma_{\phi_{2}\psi\bar{\psi}}(p',p) &\equiv \displaystyle
\frac{1}{Z_{\phi_{2}\psi\bar{\psi}}(m^2, m^2, 0)} 
\biggl[ -i\gamma_5 F_0+F_1+i\gamma_5\Pslash
F_2+\Pslash F_3 
 +i\Deltaslash(1+\theta)\frac{1+\gamma_5}{2} G_+ \\
&\displaystyle + i\Deltaslash\theta
\frac{1-\gamma_5}{2}
G_- + (H_+(1+\theta)
\frac{1+\gamma_5}{2}+H_-\theta\frac{1-\gamma_5}{2})(\Deltaslash\Pslash-\Pslash\Deltaslash) \biggr] ,
\end{array}
\eqno{(2.23)}
$$
where $F_i, G_{\pm},$ and $H_{\pm}$ are invariant amplitudes which
depend on $p^2,p'^2$ and $\Delta^2$, and $P\equiv \frac{1}{2}(p'+p)$.
Also, $F_0 (m^2, m^2, 0) = 1$ by definition.
By power counting, $F_1$ is logarithmically divergent, while $F_2,
F_3, G_{\pm}$ and $H_{\pm}$ are ultraviolet finite.  Substituting (2.23)
into (2.22), we get
$$
\frac{m_{0}Z}{Z_{\phi_2\psi\bar{\psi}}(m^2,m^2,0)}=m \bigl[ 1+a(m^2) \bigr] , 
\eqno{(2.24)}
$$
and
$$
F_i(m^2,m^2,0)=0, \hspace{.50in} i = 1,2,3.
\eqno{(2.25)}
$$

Equation (2.25) with $i=1$ insures that the amplitude $F_1$ is actually
ultraviolet convergent.  Let us define the
renormalized coupling constant $f$ by
$$
f \equiv f_0 \frac{\sqrt
{Z_{\phi_2}(0)}Z_\psi(m)}{Z_{\phi_2\psi\bar{\psi}}(m^2,m^2,0)},
\eqno{(2.26)}
$$
with $\displaystyle\frac{v_0}{\sqrt{2}}$ the bare vacuum expectation value of $\phi$.  Then we obtain from
(2.24) and (2.26) that
$$
f=\frac{m}{v}[1+a(m^2)]Z_\psi(m^2),
\eqno{(2.27)}
$$
where the renormalized vacuum expectation value $v$ is given by\footnotemark[1]
$$
v\equiv v_{0}Z/\sqrt{Z_{\phi_{2}}(0)}.
$$
By (2.15b), eq. (2.27) can be written as\footnotemark[1]
$$
f=\frac{m}{v}[1+a_r(m^2)].
\eqno{(2.28)}
$$
Thus the values of $m$ and $f$ are related.  Hence the renormalized Yukawa
coupling constant $f$ is finite if $m$ is finite, and vice versa.

Let us next study the Ward-Takahashi identity (2.20) in the limit where 
$\Delta$ is infinitesimal but not zero.  We shall keep track of terms in this
identity which are linear in $\Delta$.  We remark that the point of subtraction
for $\Gamma_{V\psi\bar{\psi}}$ requires some care.  This is because
that in the Landau gauge adopted in our formalism,
$\Gamma_{V\psi\bar{\psi}}$ is infrared divergent if we set both $p^2$ and
${p'}^2$ to $m^2$, as $\phi_2$ is massless in the Landau gauge.  A more
complete discussion on this can be found in Appendix B.  We shall only mention
here that such a divergence is superficial, as all infrared divergent terms in
a physical scattering amplitude must cancel.  The point is that there is no
infrared divergence for the physical amplitudes evaluated in the 
$\alpha$-gauge with $\alpha \ne 0$.  Since the on-shell amplitudes are $\alpha$
independent$^{6}$, the on-shell amplitudes in the Landau gauge are
infrared finite.  More precisely, at $p^2 = p'^2 = m^2$, $\bar{u}(p') 
\Gamma_{V\psi\bar{\psi}} u(p)$ is
infrared finite.  Nevertheless, some of the matrix elements of 
$\Gamma_{V\psi\bar{\psi}}$
do have infrared divergence when $p^2$ and $p'^2$ are both equal to $m^2$.
Thus the point of subtraction for $\Gamma_{V\psi\bar{\psi}}$ will be chosen 
to be at $p^2 = p'^2
= \Omega^2$, where $\Omega^2$ is not equal to $m^2$.  The physical 
amplitudes are $\Omega$-independent. 
By Lorentz covariance, $\Gamma^\mu_{{V}\psi\bar{\psi}}$ has
terms proportional to $\gamma^\mu, P^\mu$, and $\Delta^\mu$.  We shall
ignore terms proportional to the latter two in the Ward-Takahashi identity (2.20) where 
$\Gamma^\mu_{V\psi\bar{\psi}}$ is dotted with $\Delta_\mu$.  This is
because in the limit in which $\Delta$ is infinitesimal with $p^2 = p'^2 = \Omega^2$
$$
\Delta_\mu P^\mu = \frac{p'^2-p^2}{2}=0,
$$
while $\Delta_\mu\Delta^\mu$ is quadratic in $\Delta$.  Thus we shall replace
$\Gamma_{V\psi\bar{\psi}}$ in the Ward-Takahashi identity by
$$
\frac{1}{Z_{V\psi\bar{\psi}}(\Omega^2, \Omega^2, 0)} \biggl[ (\bar{\alpha}_+ +2
\beta_+ \Pslash)\gamma^\mu (1+\theta)
\frac{1+\gamma_5}{2}+(\bar{\alpha}_- +2\beta_- \Pslash)\gamma^\mu \theta
\frac{1-\gamma_5}{2} \biggr] . 
\eqno{(2.29)}
$$
By power counting, the invariant amplitudes $\beta_+$ and $\beta_-$ are
ultraviolet finite.  We shall therefore pay attention to
$Z_{V\psi\bar{\psi}}$ and $\bar{\alpha}_\pm$ only.  In the lowest order, 
$\bar{\alpha}_+$
and $\bar{\alpha}_-$ are both equal to 1.  We shall define $\bar{\alpha}_+$ to be unity
at the subtraction point.  Obviously, $\bar{\alpha}_-$ is not necessarily equal to
unity at the point of subtraction.  Substituting (2.29) and
(2.23) into (2.20) and equating the terms proportional to
$\displaystyle -i\Deltaslash(1+\theta)\frac{1+\gamma_5}{2}$, we get
$$
\frac{Z_{\psi} (m)}{Z_{V\psi\bar{\psi}}(\Omega^2,\Omega^2,0)}
= 1 - 
c_r(\Omega^2)
- d_r(\Omega^2) - m[1+a_r(m^2)]G_+(\Omega^2,\Omega^2,0).
\eqno{(2.30)}
$$
Eq. (2.30) shows that $Z_\psi / Z_{V\psi\bar{\psi}}$ is a finite number --
the counterpart of $Z_2/Z_1 = 1$ in QED.  By equating coefficients of 
$\displaystyle -i\Deltaslash \theta \frac{1-\gamma_5}{2}$ in (2.20), in the
limit $\Delta$ infinitesimal and $p^2 = p'^2 = \Omega^2$, we get
$$
\frac{Z_{\psi} (m)}{Z_{V\psi\bar{\psi}}(\Omega^2, \Omega^2, 0)} \bar{\alpha}_- (
\Omega^2, \Omega^2, 0) = 1 - c_r(\Omega^2) + d_r(\Omega^2) - m[1+a_r(m^2)]
G_-(\Omega^2, \Omega^2, 0),
\eqno{(2.31)}
$$
which shows that $\bar{\alpha}_-(\Omega^2, \Omega^2, 0)$ is finite.  This also
means that the bare parameter $\theta$ needs no renormalization.

Using the rules of subtractions we have given in this section together
with the BPHZ formalism, one may perform renormalized perturbative
calculations to all orders.  The renormalized parameters in the Feynman
rules are required to obey relations such as (2.30).  The
presence of $\gamma_5$, the meaning of which is controversial in
dimensional regularization, presents no difficulty whatsoever in our
approach.

One may also make use of the Ward-Takahashi identities for Green's functions
of bosons to show that the physical constants associated with bosons are 
finite$^1$.  Generalization to other gauge field theories such as that of
$SU(2) \times U(1) \times SU(3)$ is straightforward.
As a demonstration, we shall evaluate the lowest-order as well as the
next order triangular anomalies in the Abelian Higgs theory in the
sections below.

\mysection{The Triangular Anomaly}

\mysubsection{General Considerations}

In this section we consider the amplitude with three external gauge mesons
with polarizations $\displaystyle \mu, \nu,\rho$ and outgoing momenta $k_1,
k_2, k_3,$ respectively, where
$$
k_1+k_2+k_3=0 .
\eqno{(3.1)}
$$

We first observe that this amplitude is linearly-divergent by power
counting.  We shall show, however, that it is not only ultraviolet
finite but also has a definite value as a consequence of Bose
statistics.  Indeed, there is no undetermined constant in this
amplitude, despite the appearance of linear divergence.  This is true to all
perturbative orders.  

To see this, take, for example, the lowest-order term of this amplitude,
the diagrams for which are illustrated in Fig. 1.  We make two remarks:

(i) Only two of the six diagrams shown in Fig. 1 are topologically
independent.  However, since the amplitudes are linearly divergent, two
diagrams with different designation of loop momenta yield different
amplitudes.  Let us denote as $M(1,2,3)$ the amplitude corresponding to
diagram 1(a), where $1$, for example, denotes the first gauge meson,
with momentum $k_1$ and polarization $\mu$.  Then the amplitude in
diagram $1(e)$, with the explicit designation of the loop momentum given in
the figure,  is $M(2,1,3)$, i.e., it is obtained from $M(1,2,3)$ by
interchanging $(k_1,\mu)$ with $(k_2,\nu)$.  The amplitudes for the
other diagrams in Fig. 1 are obtained similarly.  The lowest-order
$V\!\!\!-\!\!V\!\!\!-\!\!V$ amplitude is therefore equal to

$$
 \frac{1}{3}[M(1,2,3)+M(2,3,1)+M(3,1,2)
 +M(1,3,2)+M(2,1,3)+M(3,2,1)],
\eqno{(3.2)}$$

The factor $1/3$ is combinatoric, accounting for the overcounting of
independent diagrams.  We emphasize that the amplitude in (3.2)
satisfies Bose statistics \underline{explicitly}, i.e., while $k_1, k_2$
and $k_3$ are linearly dependent, Bose symmetry is obeyed without the
need to invoke (3.1).

(ii) While the amplitude $M(1,2,3)$ is linearly divergent, it is
straightforward to show that the sum of $M(1,2,3) + M(2,1,3)$ is only
logarithmically divergent.  Thus the expression in (3.2) is
logarithmically divergent by power counting.

(iii) In order to incorporate Furry's theorem explicitly, we add to
diagram 1(a) the diagram obtained from it by reversing the direction of
the loop.  The amplitude $M(1,2,3)$ will be taken to be half
the sum of the amplitudes corresponding to these two diagrams.

With these considerations we shall put
$$\begin{array}{ll}
\Gamma^{\mu\nu\rho}_{VVV} & =
\epsilon^{\mu\nu\rho\sigma}[k_{1\sigma}W_1(k^2_1,k^2_2,k^2_3)+k_{2\sigma}W_1(k^2_2,k^2_3,k^2_1)+ k_{3\sigma}W_1(k^2_3,k^2_1,k^2_2)]\\

& +\epsilon^{\mu\nu\sigma\sigma'}k_{1\sigma}k_{2\sigma'}[k^\rho_1
W_2(k^2_1,k^2_2,k^2_3)+ k^\rho_2 W_3(k^2_1,k^2_2, k^2_3) + k^\rho_3 W_4(k^2_1,k^2_2, k^2_3)]\\

& +\epsilon^{\nu\rho\sigma\sigma'} k_{2\sigma}k_{3\sigma'} [k^\mu_2 W_2(k^2_2,
k^2_3, k^2_1) + k^\mu_3 W_3(k^2_2, k^2_3, k^2_1) + k^\mu_1 W_4(k^2_2, k^2_3, k^2_1)]\\

& + \epsilon^{\rho\mu\sigma\sigma'}k_{3\sigma}k_{1\sigma'}[k^\nu_3 W_2(k^2_3,
k^2_1, k^2_2) + k^\nu_1 W_3(k^2_3, k^2_1, k^2_2) + k^\nu_2 W_4(k^2_3, k^2_1, k^2_2)].
\end{array}$$
$$\eqno{(3.3)}$$

This form$^7$ is obtained by taking advantage of the explicit invariance
of (3.2) under the cyclic permutation of the external gauge mesons,
i.e., $1\rightarrow2, 2\rightarrow3, 3\rightarrow1$.  We note that all
terms in the amplitude are accompanied by the Levi-Civit$\acute{a}$
symbol $\displaystyle \epsilon^{\mu\nu \rho\sigma}$, as the other
invariant amplitudes vanish by an extension of Furry's theorem.

Complete symmetry under all interchanges of the external gauge mesons
requires, in addition, to require that $\Gamma^{\mu\nu\rho}_{VVV}$ is
invariant
under $1\Leftrightarrow 2$, which leads to
$$
W_1(k^2_1, k^2_2, k^2_3) = -W_1(k^2_1, k^2_3, k^2_2),
\eqno{(3.4)}
$$
$$
W_2(k^2_1, k^2_2, k^2_3) = W_3(k^2_2, k^2_1, k^2_3),
\eqno{(3.5)}
$$
and
$$
W_4(k^2_1, k^2_2, k^2_3) = W_4(k^2_2, k^2_1, k^2_3).
\eqno{(3.6)}
$$
The explicit invariance of (3.3) under the other exchanges of external
gauge mesons yields no new relations.

We shall pay particular attention to (3.4), which gives
$$
W_1(k^2_1, 0,0)=0.
\eqno{(3.7)}
$$
In particular, (3.7) gives
$$
W_1(0,0,0) = 0. 
\eqno{(3.8)}
$$
Equation (3.8) together with (3.3) imply that $\Gamma_{VVV}^{\mu\nu\rho}$
together with its partial derivatives vanish at $k_1=k_2=k_3=0$.  Since
the amplitude $\Gamma_{VVV}^{\mu\nu\rho}$ is linearly divergent by power
counting, the two conditions of subtraction at $k_1=k_2=k_3=0$ make it
finite.  As a consequence, the evaluation of the 
$V\!\!\!-\!\!V\!\!\!-\!\!V$ amplitude requires no regularization.  
In practice, we integrate over $p$ symmetrically in exactly the same way
for all of the amplitudes corresponding to the diagrams in Fig. 1.

The relation (3.8) holds not only for the lowest-order, but in arbitrary
orders.  This means that the amplitude $\Gamma^{\mu\nu\rho}_{VVV}$ is
ultraviolet finite to all orders.  We note that the Lagrangian does not
provide a counter term of the form of $V^3$.  But the amplitude
$\Gamma^{\mu\nu\rho}_{VVV}$ needs no counter term.  Therefore, the
Abelian-Higgs theory with a chiral-fermion field is renomalizable.  The
existence of anomalies, which we shall demonstrate in the rest of this
paper, is of no relevance to renormalizability.

\mysubsection{Triangular Anomaly in the Lowest Order}

Over a quarter of a century ago, Adler$^8$, Bell, and Jackiw$^9$
discovered the axial anomaly$^{10}$.  They considered the amplitude for the
triangular graph with two vector vertices and one axial-vector vertex.
While this amplitude is linearly divergent, enforcement of current
conservation on the vector vertices fixes the arbitrary constant of the
linearly-divergent integrals.  They then showed that the current
conservation on the axial-vector vertex, argued to be valid in the limit
$m\rightarrow 0$ as a consequence of chiral invariance, is not satisfied
by this amplitude.

Let us examine the considerations above in the context of the
Abelian-Higgs model, in which the vector meson couples to the chiral
fermion with the form given in (2.16).  In this model, there is no
conservation law either for the vector current or the axial vector
current.  Indeed, since the gauge vector meson couples to the chiral
fermion with the form given in (2.16), neither the vector current nor
the axial-vector current is meaningful individually.  Furthermore, the
quantization of a gauge field requires adding gauge-fixing terms and
ghost terms to the Lagrangian.  The equations of motion containing the
contribution of such terms are different from the corresponding ones
used in the classical version of the theory.  The current-conservation
laws are replaced by the Ward-Takahashi identities, which are the
consequences of the invariance of the Lagrangian under BRS
transformations.  The Ward-Takahashi identity for the
$V\!\!\!-\!\!V\!\!\!-\!\!V$ amplitude in the Landau gauge is $$
ik_{1\mu}\Gamma^{\mu\nu\rho}_{VVV}(k_1,k_2,k_3)-Z
m_0\Gamma_{\phi_{2VV}}^{\nu\rho}(k_1,k_2,k_3)=0,
\eqno{(3.9)}
$$ where $\Gamma_{VVV}(\Gamma_{\phi_2VV)}$ is defined with a factor
$g^3_0(f_0g_0^2)$ taken away from the
$V\!\!\!-\!\!V\!\!\!-\!\!V(\phi_2\!\!\!-\!\!V\!\!\!-\!\!V)$ amplitude.
The divergent amplitude $\Gamma^{\mu\nu\rho}_{VVV}$ is uniquely
determined by the requirement of Bose statistics instead of that of
vector current conservation.  As it turns out, the Ward-Takahashi
identity (3.9) is not necessarily valid.  In the Abelian Higgs model
with one chiral fermion, an additional term must be amended to the
right-hand side of (3.9).  This term is the ABJ anomaly in the Abelian
Higgs model.

We first calculate the lowest-order term for
$\Gamma_{\phi_{2}VV}^{\:\:\:\nu\rho}$ contributed by a single chiral fermion,
 the two Feynman diagrams for
which are illustrated in Fig. 2.  The lowest-order amplitude for
$\Gamma_{\phi_{2}VVV}^{\:\:\:\nu\rho}$ is equal to

$$
T^{\nu\rho}(k_1,k_2,k_3) = \int \frac{d^4p}{(2\pi)^4}
\frac{N_a+N_b}{(p^2-m^2)[(p-k_2)^2-m^2][(p+k_1)^2-m^2]},
\eqno{(3.10)}
$$
where
$$
N_a \equiv Tr \biggl[ \gamma_5(\pslash+\kslash_1+m)\gamma^\rho
\frac{1+2\theta+\gamma_5}{2}(\pslash-\kslash_2+m)\gamma^\nu
\frac{1+2\theta+\gamma_5}{2}(\pslash+m) \biggr] ,
\eqno{(3.11)}
$$
and
$$
N_b \equiv Tr \biggl[ \gamma_5(-\pslash+m)\gamma^\nu
\frac{1+2\theta+\gamma_5}{2}(-\pslash+\kslash_2+m)\gamma^\rho
\frac{1+2\theta+\gamma_5}{2}(-\pslash-\kslash_1+m) \biggr] .
\eqno{(3.12)}
$$
After some algebra, we
get the lowest-order term for $\Gamma^{\nu\rho}_{\phi_2VV}$ as
$$
-\frac{m}{4\pi^2}[(1+\theta)^3 - \theta^3]\epsilon^{\nu\rho\sigma\sigma'}k_{1\sigma}k_{2\sigma'}\int
\frac{d^3\alpha\delta(1-\Sigma\alpha)(1-\alpha_1)}{m^2-\alpha_2\alpha_3
k^2_1-\alpha_3\alpha_1 k^2_2-\alpha_1\alpha_2 k^2_3} ,
\eqno{(3.13)}
$$
where we have symmetrized with respect to $\alpha_2, \alpha_3$ and $k_2, k_3$.

Next we turn to the lowest-order term in the renormalized perturbation
series for $\Gamma^{\mu\nu\rho}_{VVV}$.  
It is convenient to define
$$
L \equiv \frac{1}{2} (1+\gamma_5) {\hskip 0.5cm} \mbox{and} {\hskip 0.5cm}
R \equiv \frac{1}{2} (1-\gamma_5) .
\eqno(3.14)
$$
We then have
$$
L^2 = L ,
\eqno(3.15)
$$
$$
R^2 = R ,
\eqno(3.16)
$$
$$
LR = RL = 0,
\eqno(3.17)
$$
and
$$
\frac{(1+2\theta+\gamma_5)}{2} = (1+\theta) L + \theta R .
\eqno(3.18)
$$ Note that $L$ is the projection operator for a left-handed fermion
and similarly for $R$.  A left-handed fermion remains to be
left-handed after interacting with a gauge vector meson, and so does
the right-handed fermion.  Therefore, a fermion in the loop of the
diagrams in Fig. 1 is either left-handed or right-handed throughout
the loop.  Consequently, we may calculate the contributions from the
left-handed fermion and the right-handed fermion separately.  The
contributions from the left-handed fermion is equal to that from a
pure left-handed fermion $(\theta = 0)$ multiplied by a factor of
$(1+\theta)^3$.  This is because there are three gauge vector meson
vertices on the fermion loop, with each vertex associated with a
factor of $(1+\theta)$.  Similarly, the contributions from the
right-handed fermion is equal to that from a pure right-handed fermion
multiplied by a factor of $\theta^3$.  Consequently, the term in
$\Gamma^{\mu\nu\rho}_{VVV}$ corresponding to diagram (1a) is given by

$$
-\frac{i}{2} [(1+\theta)^3 - \theta^3]\int \frac{d^4p}{(2\pi)^4}
\frac{Tr[\gamma_5\gamma^\mu(\pslash+\kslash_1)\gamma^\rho(\pslash-\kslash_2)\gamma^\nu
\pslash]} {[(p-k_2)^2-m^2][(p+k_1)^2-m^2](p^2-m^2)} .
\eqno{(3.19)}
$$
In obtaining (3.19),
 we have averaged over the two directions of the loop momentum.  It is only
after this is done that the argument of the trace in (3.19) has exactly
one factor of $\gamma_5$.  

The evaluation of (3.19) is straightforward.  We find that the lowest-order
amplitude is in the form (3.3), with
$$
{W^{(0)}_i(k^2_1,k^2_2, k^2_3)} = \frac{i}{12\pi^2} \int \frac{[ (1+\theta)^3 - \theta^3] N_i}{m^2-\alpha_2\alpha_3k^2_1-\alpha_3\alpha_1k^2_2-\alpha_1\alpha_2k^2_3}
d^3\alpha\delta(1-\sum\alpha) \hspace{.25in} [i =1,2,3,4]
\eqno{(3.20)}$$
with
$$
\begin{array}{lll}
N_1& = &
2(\alpha_2-\alpha_3)\alpha_2\alpha_3k^2_1
+
\alpha_3(3\alpha_1\alpha_2+\alpha_1\alpha_3+\alpha^2_2+\alpha_2\alpha_3)k^2_2\\
&&
\null - 
\alpha_2(3\alpha_3\alpha_1+\alpha_1\alpha_2+\alpha_2\alpha_3+\alpha^2_3)k^2_3,\\
N_2& = & - 2\alpha_2\alpha_3,\\
N_3& = & - 2\alpha_1\alpha_3,\\
N_4& = &   4\alpha_1\alpha_2,\\
\end{array}
$$
where $W^{(0)}_i$ is the lowest-order amplitude for $W_i$ defined in
(3.3).  From (3.20) and (3.13), it is also easy to find that the right
side of (3.9) should be amended with an anomalous term, the lowest-order
of which is
$$
\epsilon^{\nu\rho\sigma\sigma'}k_{1\sigma}k_{2\sigma'}\frac{1}{12\pi^2}
[ (1+\theta)^3 - \theta^3 ],
\eqno{(3.21)}
$$
in the Abelian Higgs theory with one chiral fermion field.

\mysection{Anomaly of the Next Order}

\mysubsection{General Considerations}

It is customary to introduce a family of chiral fermions with
appropriate quantum numbers in such a way that the sum of their
contributions to the lowest-order anomaly vanishes.  With this being
done, let us go on to study the anomaly in the next order.

In the next two sections of this paper, we shall calculate the
radiative corrections of the triangular anomaly in the Abelian Higgs
theory from a single chiral fermion.  The contributions from the family of
chiral fermions can be obtained by adding the contributions from each of the
chiral fermions.  As we shall see, this former anomaly, 
expressed in terms of
renormalized parameters, is equal to zero$^{12}$.

Before we plunge into the details of the calculations, a few words of
perspective may be in order.  The anomaly amplitude is a Feynman
integral.  Because of the symmetry of the Lagrangian, the integrand
for the anomaly amplitude consists of terms which, were we allowed to
change freely the designation of loop momenta for each of them, would
cancel completely.  
However, the anomaly amplitude is a linearly divergent integral, and
shifting the variable of integration of such an integral without
compensation is not legitimate.  We shall call such illegitimate
cancellations superficial cancellations.  In the lowest-order anomaly
amplitude, there is only one loop-momentum, and the amount of shifts
required for cancellation is a linear superposition of the external
momenta.  Consequently, the invariant amplitude of the lowest-order
anomaly can be expressed as an integral over a surface on which the
loop momentum is infinite, as commonly done in the literature.  As a
result, the anomaly amplitude is a constant, as the external momenta
which appear in the integrand of this amplitude can be dropped.  In
the anomaly amplitude of the next lowest-order, there are two
loop-momenta, and the shift for one of the loop-momentum required for
cancellation may involve the other loop-momentum.  Nevertheless, since
the Jacobian of integration for such a shift is equal to unity, the
contributions to the anomaly amplitude again come from a region in
which the loop momenta are infinitely large.  In this region, the
external momenta are again negligible, hence the anomaly amplitude in
the next lowest-order is also a constant.

\mysubsection{Reduction of Amplitude}

The relevant diagrams for $\Gamma^{\mu\nu\rho}_{VVV}$ in the lowest and
next to the lowest orders are illustrated in Fig. 3 and Fig. 4.  The
diagrams in Fig. 3 are just the lowest-order triangular diagrams with insertions of
radiative corrections to the vertices and internal lines.  Instead of
using six diagrams of this kind with permuted external lines, (and
dividing the sum of the corresponding amplitudes by 3), we shall only use
two diagrams here.  Bose symmetry is maintained by an ingenious choice of
the loop-momenta$^{13}$.  Referring to Fig. 3, we see that, with the
loop-momentum so chosen, diagram (3a) and diagram (3b) are individually
invariant under the cyclic permutation of $1\rightarrow2, 2\rightarrow3,
3\rightarrow 1$.  In addition, these two diagrams turn into each other
under $2\leftrightarrow3$.  Therefore, with the understanding that the
integration over $p$ is symmetrically performed in exactly the same way,
 the sum of the
amplitudes corresponding to these two diagrams is completely
Bose-symmetric.

Next we turn to the diagrams in Fig. 4.  We note that each of the first
two diagrams in Fig. 4 has four internal fermion lines, while each of the
rest of the diagrams has three internal fermion lines.  The diagrams
to be included in our calculations for $\Gamma^{\mu\nu\rho}_{VVV}$ are
obtained from diagrams (4a)--(4f) by permuting the external lines in all
possible ways.

The diagrams relevant to the amplitude $\Gamma_{\phi_{2}VV}$ are
obtained from those for the amplitude $\Gamma_{VVV}$ by replacing an
external $V$ line by an external $\phi_2$ line, plus the diagrams obtained
from diagram (4g) by permuting the external lines in all possible ways.  

Since there are a large number of diagrams for each of the amplitudes
$\Gamma_{VVV}$ and $\Gamma_{\phi_{2}VV}$, the calculations of these
individual amplitudes are tedious.  However, it is much simpler to
calculate the anomaly amplitude contributed by these diagrams, as we may
take advantage of the Ward-Takahashi identities such as (2.20).

Let us first consider the anomaly amplitude (the expression we should add to
the right-hand side of
(3.9)) contributed by the diagrams in Fig. 3.  More precisely, we
calculate $\Gamma_{VVV}$ contributed by such diagrams and
$\Gamma_{\phi_{2}VV}$ contributed by those obtained by replacing the
gauge meson of momentum $k_1$ with a $\phi_2$ meson in these diagrams.
By (2.17), the anomaly amplitude from diagram (3a) is:
$$\begin{array}{ll}
&- \displaystyle \int {\frac{d^4p}{(2\pi)^4} }
Tr \biggl[ \Gamma^\nu(p-k_{23},p-k_{12})\frac{1+2\theta+\gamma_5}{2}
S(p-k_{31})\Gamma^\rho(p-k_{31},p-k_{23})S(p-k_{23}) \biggr] \\
& + \displaystyle \int {\frac{d^4p}{(2\pi)^4} }
Tr \biggl[ \Gamma^\nu(p-k_{23},p-k_{12})S(p-k_{12})\frac{1+2\theta-\gamma_5}{2}\Gamma^\rho(p-k_{31},p-k_{23})S(p-k_{23}) \biggr] ,
\end{array}
\eqno{(4.1)}
$$
and the anomaly amplitude from diagram (3b) is
$$\begin{array}{ll}
&- \displaystyle \int {\frac{d^4p}{(2\pi)^4} } Tr \biggl[ \Gamma^\nu(p+k_{12},
p+k_{23})S(p+k_{23})\Gamma^\rho(p+k_{23},
p+k_{31})\frac{1+2\theta +\gamma_5}{2}S(p+k_{12}) \biggr] \\
&+ \displaystyle \int {\frac{d^4p}{(2\pi)^4} } Tr \biggl[ \Gamma^\nu(p+k_{12},
p+k_{23})S(p+k_{23})\Gamma^\rho(p+k_{23},
p+k_{31})S(p+k_{31})\frac{1+2\theta-\gamma_5}{2} \biggr] .
\end{array}
\eqno{(4.2)}
$$
In the above
$$
k_{ij}\equiv(k_i-k_j)/3.
$$
The four terms in (4.1) and (4.2) are schematically represented by the
four diagrams in Fig. 5.  We note that in these diagrams, the dotted
line located behind (in front of) the vertex carries a factor
$\displaystyle \frac{1+2\theta+\gamma_5}{2} (\frac{1+2\theta -\gamma_5}{2})$, the order
being dictated by the direction of the arrow.  (See the right-hand side
of (2.20) for the origin of the factors $(1+2\theta\pm\gamma_5)/2)$.  We also
mention that the momentum variables associated with a vertex which is
joined by a dotted line are specified by the momentum of the external
gauge meson and the momentum of the fermion line on the opposite side of
the dotted line.  For example, for the vertex function $\Gamma^\nu$ in
diagram (5a), the outgoing momentum is $p-k_{23}$, while the incoming
momentum is not $p-k_{31}$, but is
$$
(p-k_{23})+k_2=p-k_{12}.  
$$
Let us change the loop momenta in diagram (5a) and in diagram (5d) so
that the momenta for both the upper fermion lines are designated as $p$.
Similarly, we change the variable of integration for diagrams (5b) and
(5c) so that the momenta for both lower fermion lines in these diagrams
become $p$.  The anomaly amplitude given by the sum of (4.1) and (4.2)
then becomes
$$
J_1 + J_2
\eqno(4.3)
$$
where $J_{1}$ is a volume integral given by
$$
J_1 = \int \frac {d^{4} p} {(2\pi)^4}
Tr (N_{5}+N_{6})
\eqno(4.4a)
$$
with
$$\begin{array}{ll}
N_5 = \biggl[ & \displaystyle -\Gamma^\nu(p+k_3,
p-k_1)\frac{1}{2}(1+2\theta+\gamma_5)+\frac{1}{2}(1+2\theta-\gamma_5)\Gamma^\nu(p-k_2,
p) \,\, \biggr] \\
& S(p)\Gamma^\rho(p,p+k_3)S(p+k_3) , \\
\end{array}
\eqno{(4.4b)}
$$
and
$$
N_6={\Gamma^\nu(p,p+k_2)S(p+k_2) \biggl[ -\Gamma^\rho(p+k_2, p-k_1)\frac{1+2\theta+\gamma_5}{2}+\frac{1+2\theta-\gamma_5}{2}\Gamma^\rho(p-k_3,p) \biggr] S(p).}
\eqno{(4.4c)}
$$
The term $J_2$ comes from changing the variable of integration for the
linearly divergent integrals, and is given by the surface integral 
$$
J_2
\equiv
\frac{1}{3}
\int \frac{dS^\mu}{(2\pi)^4} N_{7},
\eqno{(4.5a)}
$$
where $N_{7}$ is given by
$$
\begin{array}{lll}
N_{7}&=& \displaystyle
(k_3-k_1)_\mu Tr[\Gamma^\nu (p+k_3,
p-k_1)\frac{(1+2\theta+\gamma_5)}{2} S(p)\Gamma^\rho(p,p+k_3)S(p+k_3)]\\ 
&+&\displaystyle
(k_2-k_3)_\mu Tr[\frac{(1+2\theta-\gamma_5)}{2}\Gamma^\nu(p-k_2,p)S(p)\Gamma^\rho(p,p+k_3)S(p+k_3)]\\
&-&\displaystyle
(k_1-k_2)_\mu
Tr[\Gamma^\nu(p,p+k_2)S(p+k_2)\Gamma^\rho(p+k_2,p-k_1)\frac{(1+2\theta+\gamma_5)}{2}S(p)]\\
&-&\displaystyle
(k_2-k_3)_\mu Tr[\Gamma^\nu((p,p+k_2)S(p+k_2)\frac{(1+2\theta-\gamma_5)}{2}\Gamma^\rho(p-k_3,
p)S(p)] ,
\end{array}
\eqno(4.5b)$$
and where the integration is over a three-dimensional surface in
the p-space at infinity.

Since we are interested in the anomaly only in the next lowest order,
(4.4b) and (4.4c) can be replaced by
$$\begin{array}{ll}
N_5=
\biggl[ & \displaystyle \Gamma^\nu(p+k_3, p-k_1)
\frac{(1+2\theta+\gamma_5)}{2}
-
\frac{(1+2\theta-\gamma_5)}{2}
\Gamma^\nu(p-k_2,p) \biggr] \\
& \displaystyle \cdot \frac{1}{\pslash-m}
\gamma^\rho \frac{(1+2\theta+\gamma_5)}{2} \frac{1}{\pslash+\kslash_3-m} ,
\end{array}
\eqno(4.6a)
$$
and
$$\begin{array}{ll}
\displaystyle N_6=\gamma^\nu
\frac{(1+2\theta+\gamma_5)}{2} \frac{1}{\pslash+\kslash_2-m} \biggl[ &\displaystyle \Gamma^\rho(p+k_2,p-k_1)\frac{(1+2\theta+\gamma_5)}{2} \\
& \displaystyle -\frac{(1+2\theta-\gamma_5)}{2}\Gamma^\rho(p-k_3,p) \biggr] \frac{1}{\pslash-m} ,
\end{array}
\eqno{(4.6b)}
$$
as the difference of the two vertex functions inside each of the
brackets is already at least of that order.

The vertex function $\Gamma$ in (4.4), (4.5) and (4.6) is that of the
$V\psi\bar{\psi}$-vertex.  We shall express this vertex function in a
renormalized perturbation series.  Since the counter term for this
vertex function is equal to a constant times
$\displaystyle \gamma^\mu\frac{(1+2\theta+\gamma_5)}{2}$, it is easy to
prove that  
the counter terms for the $\Gamma$ function in (4.6) cancel.

We draw in \fig6 diagrams for the lowest order radiative correction
of $\Gamma_{\psiV}$.
In particular, diagram (6a) depicts the radiative
correction to $\Gamma_{\psiV}$ due to the exchange of a vector meson.
We draw in \fig7 schematic diagrams representing $J_1$ of
(4.4), with the diagram~(6a) inserted for the vertex sub-graph.  Note
that the dotted lines in \fig7 are always joined to bare
 $\psiV$ vertices, each of which is assigned a factor
 $\displaystyle \gamma^\mu [(1+\theta) L + \theta R]$ in the Feynman rules.
It is easy to see that the amplitude corresponding to diagram~(7a) is
equal to that corresponding to diagram~(7d) after the change of
integration variables
$$
p\rightarrow p-q-k_2
$$
is made.  This does not mean, however, that these two divergent
amplitudes cancel each other, as a change of integration variables is
not allowed.  Similarly, the anomaly amplitude corresponding to
diagram~(7b) is equal to that corresponding to diagram~(7c), after a
change of variable is made.  The sum of the amplitudes corresponding to
these four diagrams will be evaluated in the next section.

Next we consider the anomaly amplitudes corresponding to $J_1$ of (4.4) with 
diagram (6b) inserted for the vertex.  These anomaly amplitudes are represented
by the diagrams in Fig.8.  We emphasis two points:
(1) The integration over $q$, the loop momentum for the radiative correction of
the vertex, is to be performed before that over $p$, the loop momentum in the
triangular diagrams in Fig.3.  This order of integration must be followed strictly.  This is because individual Feynman integrals are divergent and the value
of an individual integral depends on the order of integration.  (2) The 
integration over $q$ for the sum of the anomaly amplitudes corresponding to the
diagrams in Fig.9 is convergent.  This is because the ultraviolet divergent in
the $q$-integration for diagram (8a) cancels that for diagram (8b).  Similarly
for the ultraviolet divergences in the $q$-integration for diagrams (8c) and 
(8d).  Therefore, we are allowed to choose to symmetrically integrate over $q$ for all of these
amplitudes.

The anomaly amplitudes represented by the diagrams in Fig.8 are cancelled by 
corresponding terms in the non-planar diagrams of the form of Fig.(4a).  To
see this, we draw all twelve of these non-planar diagrams in Fig.9.  Notice that
only six of them are topologically independent.  For example, diagram (9a) is
topologically the same as diagram (9h).  We draw all twelve of them for the
sake of symmetry and, to compensate for overcounting, we shall multiply the
amplitude for each of these diagrams by a factor of 1/2.  Notice also that
we have assigned the loop momenta in these diagrams in such a way that the
sum of amplitudes from these six diagrams are Bose-symmetric with respect to
the exchange of any two of external vector mesons.  The Feynman integral for
each diagram contains two integrations, namely $q$ and $p$.  We shall choose 
to integrate over $q$ first.  (The result is independent of the order of
integration, as long as we follow the same order for all the diagrams in 
Fig. 9.)  We also mention that, 
power counting, the integration over $q$ is convergent for the amplitude of
an individual diagram.  We shall therefore choose to integrate symmetrically 
over $q$.  The subsequent integration over $p$ will also be chosen to be
symmetric, as is discussed at the end of Sec.3A as well as in Sec.5.
The contribution of diagrams (9a)  and (9b) to the anomaly amplitude is
schematically evaluated in Fig.10 and Fig.11.  A dot next to the external
line of vector meson 1 in these diagrams represents the multiplication of 
$i k_{1\mu}$ to the $VVV$ amplitude.  The anomaly amplitudes are
represented by the diagrams following the equality signs.  The
diagrammatic equality is obtained via the Ward-Takahashi identity
$$
i k_\mu \Gamma^{\mu(0)}_{V \Phi_{2} H}
-
2 \mu^2_0 \Gamma^{(0)}_{\Phi_{2} \Phi_{2} H}
=
i\left[
  D^{(0)}_{H}
\right]^{-1}
-
i
\left[
    D^{(0)}_{\Phi_{2}}
\right]^{-1}
.
\eqno{(4.7)}
$$ where the superscript $(0)$ on a symbol signifies that the quantity
the symbol represents is that of the lowest order.  It is easy to see that
the sum of the anomaly amplitudes from diagram (10a) and diagram (8b) is 
equal to the anomaly amplitude from diagram (8b) with the $\gamma_5$
associated with the dotted line in diagram (8b) deleted.  Similarly,
adding the anomaly amplitudes from diagrams (10c), (11a) and (11c) simply
eliminates the $\gamma_5$ associated with the dotted lines of diagrams (8a),
(8d) and (8c) respectively.  It is also straightforward to show that the
sum of these anomaly amplitudes is equal to zero.  Briefly, the anomaly 
amplitude from diagram (8a), with the aforementioned $\gamma_5$ deleted, is
actually a convergent integral.  Therefore, it is legitimate to change the
designation of the loop momenta and this anomaly amplitude cancels the 
corresponding one from diagram (8d).  Similarly, the anomaly amplitudes
from diagrams (8b) and (8c), with the aforementioned $\gamma_5$ deleted, 
cancel each other.

We may consider in the same way the anomaly amplitudes from the diagrams
obtained from those in Fig.8 with the $\phi_2$-lines replaced by the
$H$-lines.  These anomaly amplitudes are cancelled by the anomaly
amplitudes from diagrams (10b), (10d), (11b) and (11d).

Next we consider the diagrams in Fig. 5 with the diagrams (6f) and (6g)
inserted for the vertex.  The resulting diagrams are illustrated in Fig. 12.
We note that two of the external lines attached to the triangular fermion loops
in these diagrams are scalar-meson lines.  It is therefore tempting to
argue that, since the amplitude corresponding to such a triangular
loop vanishes as a consequence of Lorentz invariance, the anomaly
amplitude corresponding to a diagram with such a loop also vanishes.
Such an argument is incorrect, as we do not integrate over the 
momentum of the fermion loop first.  

We draw in Fig. 13 some of the anomaly diagrams derived from the Feynman diagrams
(9c)-(9f) and (9i)-(9l).  Note that these anomaly diagrams are topologically
equivalent to the anomaly diagrams in Fig. 12.  The corresponding amplitudes
differ, however, as the loop momenta in these diagrams differ.  We shall
show in Appendix A that those diagrams in Fig. 12 gives zero contribution 
to the anomaly
amplitudes while those in Fig. 13 give nonzero contribution.  There are 
also other Feynman diagrams which can be reduced to diagrams as those shown
in Fig. 12 which we do not show here since they also give zero contribution
to the anomaly amplitude .  The sum of the anomaly amplitudes from the
diagrams in Fig. 13 is equal to 
$$
\displaystyle
J_3 = \epsilon^{\nu \rho \sigma \sigma'} \frac{k_{1 \sigma} k_{2 \sigma'}}
{64 \pi^4} f^2 \theta (1 + \theta) .
\eqno(4.8)
$$

Next we consider the anomaly amplitude corresponding to the diagrams in
\fig7 with diagram (6d) replacing diagram (6a) as the vertex subgraph.
It is straightforward to see that these anomaly amplitudes are exactly
canceled by the anomaly amplitudes generated by diagram (4b) with the
vector meson of momentum $k_1$
attached to the fermion loop, and diagram (4d) with the
vector meson of momentum $k_1$
at $V{\phi_2}H$ vertex.  To demonstrate this, we draw
in Fig. 14 a set of such anomaly amplitudes which exactly cancel.
Diagram (14a) is diagram (7a) with diagram (6d) replacing (6a) as the
vertex subgraph.  Diagram (14b) is an anomaly diagram derived from
diagram (4b).  Diagram (14c) is an anomaly diagram derived from diagram
(4d).  The factors associated with the dotted lines are
$\displaystyle \frac{(1+2\theta+\gamma_5)}{2}$, $-\displaystyle \frac{(1+2\theta
-\gamma_5)}{2}$, and $(-i)(-i\gamma_5)$,
respectively.  The sum of these three terms is zero.

In summary, the anomaly amplitudes from various diagrams generally cancel as a
consequence of the Ward-Takahashi identities.  There are four exceptions.
The first exception is the contribution from the nonplanar diagrams Figs. 
(9c), (9f), (9i) and (9l)  and is given by $J_3$.
The second exception is the volume integral $J_1$ 
corresponding to the diagrams in \fig7, for which changing
integration variables to effect cancellation is illegitimate, as we have
discussed.  The third exception is the surface integral $J_2$ given by (4.5).  
$J_1$ and $J_2$ will be evaluated explicitly in Appendix E.  Referring to
(4.8), (E.20), (E.30) and (E.31), we obtain for the sum of the above three exceptions$^{14}$
$$
\displaystyle
J_1 + J_2 + J_3 = 0. 
\eqno(4.9)
$$
The last 
exceptions are the anomaly amplitudes derived from the diagrams
(4c) and (4e).  In each of these diagrams, there is a triangular loop
with three vector meson lines attached to it.  The anomaly amplitudes
due to this triangular loop have been calculated in Section~3B.  This 
anomaly amplitude vanishes as we sum over the contributions from the family 
of chiral fermions with appropriate quantum numbers.

\mysection {Summary}

In this paper, we have demonstrated how to obtain renormalized Ward-Takahashi
identities for the Abelian-Higgs theory with chiral fermions.  With the aid
of this set of identities, one can calculate
unambiguously physical quantities by the classic method of renormalization on
the basis of subtractions.  As an example, we calculated the next order
triangle anomaly in this theory which is shown to vanish$^{14}$.  This calculation
is performed without the introduction of regularization.  Some of the 
calculations on higher order triangle anomaly in the past are based on the 
introduction
of cutoffs in the Feynman propagators  
and thus the result of such a calculation depends critically on the
method of regularization.  For example, it is well-known that shifting the
loop-momentum of a linearly divergent 
Feynman integral must be compensated by the addition of a finite and
nonzero constant.  However, if we regularize the propagators in this
Feynman integral, no such compensation is necessary for such a shift.
More precisely, $$
\begin{array}{ll}
& \displaystyle \lim_{\Lambda \rightarrow \infty} 
\int \frac{d^4p}{(2\pi)^4} \biggl\{ \frac{(p+a)_\mu}{[(p+a)^2 - m^2]^2}
\frac{\Lambda}{\Lambda - (p+a)^2} - \frac{p_\mu}{(p^2 - m^2)^2}
\frac{\Lambda}{\Lambda - p^2} \biggr\} \\ 
\ne & \displaystyle \int \frac{d^4p}{(2\pi)^4} \lim_{\Lambda \rightarrow \infty}
 \biggl\{ \frac{(p+a)_\mu}{[(p+a)^2 - m^2]^2}
\frac{\Lambda}{\Lambda - (p+a)^2} - \frac{p_\mu}{(p^2 - m^2)^2}
\frac{\Lambda}{\Lambda - p^2} \biggr\} \,\, . \\ 
\end{array}
\eqno(5.1)
$$
The left-hand side of the above equation is equal to zero as each of the
two integrals on the left-hand side is convergent when $\Lambda$ is finite.
On the other hand, each of the two terms in the integrand of the right-hand
side, with $\Lambda$ set to infinity, is the integrand of a linearly
divergent integral, and hence the right-hand side is not equal to zero.
Generally, caution must be exercised when the integrand of an integral 
depends on a cutoff parameter.  The limit of such an integral is not
always equal to the integral of the limit of the integrand as the cutoff
parameter is made to go to infinity.  This means that the value of a
divergent integral is dependent on the particular method of regularization.

In contrast, we start with a rigorous theory on the basis of a gauge
invariant Lagrangian with the symmetry of the vacuum spontaneously
broken.  We canonically quantize this theory, adding gauge fixing
terms and introducing the associated ghost fields.  The ghost terms
are so constructed that the Lagrangian has the BRST invariance.  As a
consequence of the BRST invariance, a system of Ward-Takahashi
identities is satisfied by the Green's functions of this theory.
While the quantum theory so obtained is ultraviolet divergent, the
Ward-Takahashi identities are taken to form the guiding rules of
renormalization.  We obtain the value of the anomaly amplitude via a
rigorous and unambiguous scheme of renormalized perturbation.  There
is no need to make up an axial vector current which is not in the
theory.  In our calculation of the anomaly amplitude, no
regularization is needed.  While a $VVV$-amplitude for an individual
Feynman diagram can be linearly divergent, the full Bose symmetry with respect to all the three external
vector mesons makes the $VVV$-amplitude ultraviolet finite without the need of
introducing a parameter of regularization.  The crucial fact is that (3.3) and
(3.8) enforce $\Gamma_{VVV}^{\mu\nu\rho}$ as well as its partial derivatives to vanish at the point where
all external momenta are zero.  One can therefore make a subtraction
at this point and the anomaly amplitude is then explicitly utlraviolet
convergent.

In practice, we choose the loop momenta for the diagrams in such a way
that full Bose symmetry is explicitly satisfied, as was illustrated,
for example, in the diagrams in Fig. 3.  We obtain the anomaly
amplitude by integrating over such loop momenta with subtractions.
Since the subtracted integral is ultraviolet finite, it is
mathematically rigorous to adopt symmetry integration for the
subtracted integral under which both the subtracted terms and the
unsubtracted terms turn out to be finite.  Because the subtracted term
satisfies the full Bose statistics explicitly on its own, it is equal to zero.
The anomaly amplitude is therefore equal to the unsubtracted term
integrated symmetrically with the loop momenta chosen as in Fig. 3 and in 
this case is found to vanish.
 
The renormalization procedure in the above can easily be 
extended to nonabelian Higgs theories with chiral fermions such as the
standard model.  The renormalized Ward-Takahashi identities so obtained 
can be used to obtain exact relations among various physical quantities that  
are of experimental interests.  Thus one can check the validity of the standard 
model against precision measurements that have been done or will be done in the
near future.

\newpage

\begin{center}
\large\bf {Appendix A}
\end{center}

In this Appendix, we calculate the anomaly amplitude corresponding to the
diagrams in Figs. 12-13.

All of the anomaly amplitudes from the diagrams in Fig. 12 are equal to
zero.  To show this, let us remind ourselves that we are going to
symmetrically integrate over $q$ first and then symmetrically integrate over
$p$.  To illustrate how this works, it is sufficient to only look at diagram 
(12a).  The Feynman integral for (12a) is
$$
\displaystyle
i \theta (1+\theta) f^2 \int \frac{d^4 p}{(2 \pi)^4} \frac{d^4 q}{(2 \pi)^4}
\frac{\cal N}{\cal D} ,
\eqno(A.1)
$$
where
$$\begin{array}{ll}
{\cal N} & = (2 q + k_2)^{\nu} Tr \biggl[ \gamma_5 \gamma^\rho (\pslash +
\kslash_3) (\pslash + \qslash - \kslash_1) \pslash \biggr] \\
& = 4 i \epsilon^{\rho\alpha\beta\sigma} p_\alpha k_{3\beta} (q - k_1)_\sigma
(2 q + k_2)^\nu .
\end{array}
\eqno(A.2)
$$
$$
{\cal D} = [(q+k_2)^2 - 2\mu^2] q^2 [(p+k_3)^2 - m^2][p^2 - m^2] [ (p+q-k_1)^2
- m^2] .
\eqno(A.3)
$$
It turns out that the anomaly amplitude given by (A.1) is independent of 
the masses in the theory.  To simplify the presentation, we shall drop the
mass terms in (A.3).  Next we introduce Feynman parameters and reduce (A.1)
into
$$
\displaystyle
24 i \theta (1+\theta) f^2 \int d^5 \alpha \delta (1- \sum \alpha) \int 
\frac{d^4 p}{(2\pi)^4} \frac{d^4 q}{(2\pi)^4} \frac{{\cal N}}{D} ,
\eqno(A.4)
$$
where
$$\begin{array}{ll}
D & = \alpha_1 (q+k_2)^2 + \alpha_2 q^2 + \alpha_3 (p+k_3)^2 + \alpha_4 p^2
+ \alpha_5 (p+q- k_1)^2 \\
& \displaystyle = (\alpha_1 + \alpha_2 + \alpha_5) (q - \delta q)^2
+ \frac{F(p)}{(\alpha_1+\alpha_2+\alpha_5)} ,
\end{array}
\eqno(A.5)
$$
where
$$
\delta q = \frac{\alpha_5 ( k_1 - p) - \alpha_1 k_2}{(\alpha_1+\alpha_2
+\alpha_5)} ,
\eqno(A.6)
$$
$$
F(p) = r (p - \delta p)^2 + (\alpha_1 + \alpha_2 +\alpha_5) {\cal A} ,
\eqno(A.7)
$$
$$
\delta p = \frac{1}{r} [ \alpha_5(\alpha_1+\alpha_2) k_1 + \alpha_1 \alpha_5
k_2 - \alpha_3 (\alpha_1+\alpha_2+\alpha_5) k_3 ] ,
\eqno(A.8)
$$
$$
r = \alpha_5(\alpha_1+\alpha_2+\alpha_3+\alpha_4) + (\alpha_1+\alpha_2)
(\alpha_3+\alpha_4) ,
\eqno(A.9)
$$
$$
{\cal A} = \frac{1}{r} \biggl\{
\alpha_2 \alpha_4 \alpha_5 k_1^2 + [ \alpha_2 \alpha_3 \alpha_5 + 
\alpha_1 \alpha_2 (\alpha_3+\alpha_4+\alpha_5) ] k_2^2 + [\alpha_1 \alpha_4
\alpha_5 + \alpha_3 \alpha_4 (\alpha_1+\alpha_2+\alpha_5) ] k_3^2 \biggr\} .
\eqno(A.10)
$$
Since the integration over $q$ is convergent, 
we make a shift in $q \rightarrow \bar{q} + \delta q$ and then integrate
over $\bar{q}$.  Upon shifting $q$, the numerator $\cal N$ takes the form
$$
{\cal N} = 2 i \epsilon^{\nu\alpha\rho\beta} \bar{q}^2 p_\alpha k_{3\beta}
+ 4 i \epsilon^{\rho\alpha\beta\sigma} (\delta q - k_1)_\alpha p_\beta
k_{3\sigma} (\delta q + k_2)^\nu .
\eqno(A.11)
$$
(A.4) becomes, after integrating over $\bar{q}$
$$
- \frac{3\theta(1+\theta) f^2}{\pi^2} \int d^5 \alpha \delta (1-\sum \alpha)
\int \frac{d^4 p}{(2\pi)^4} \biggl\{ \frac{N_1}{(\alpha_1+\alpha_2+\alpha_5)
F^2 (p)} + \frac{(\alpha_1+\alpha_2+\alpha_5) N_2}{F^3(p)} \biggr\} ,
\eqno(A.12)
$$
where
$$
N_1 = 2 i \epsilon^{\nu\alpha\rho\beta} p_\alpha k_{3\beta} ,
\eqno(A.13)
$$
$$
N_2 = 4 i \epsilon^{\rho\alpha\beta\sigma} (\delta q - k_1)_\alpha p_\beta
k_{3\sigma} (2 \delta q + k_2)^\nu .
\eqno(A.14)
$$
Now we write out explicitly $\delta q$ in terms of $p$ and make a shift in
$p \rightarrow \bar{p} + \delta p$.  Here, one should be careful about 
shifting $p$ in (A.13) since the it involves a linearly divergent integral.
Thus, when we integrate out the first term
in (A.12), it contains two pieces, the first piece is a volume integral
over $p$ and
the second piece is a constant due to shifting $p$ by an amount $\delta p$. 
The second piece is equal to
$$
\frac{3 \theta (1+\theta) f^2 \epsilon^{\nu\alpha\rho\beta}k_{1\alpha} 
k_{3\beta}} {16 \pi^4} \int \frac{d^5 \alpha \, \delta(1-\sum\alpha)}
{(\alpha_1+\alpha_2+\alpha_5)} \frac{\alpha_2 \alpha_5}{r^3} .
\eqno(A.15)
$$
The first piece is equal to (A.12) with $N_1$ replaced by 
$$
2 i \epsilon^{\nu\alpha\rho\beta} \frac{\alpha_2 \alpha_5}{r} k_{1\alpha}
k_{3\beta} .
\eqno(A.16)
$$
and $N_2$ replaced by
$$
 - 2 i \epsilon^{\nu\alpha\rho\beta} \frac{\alpha_2\alpha_5}{(\alpha_1
+\alpha_2+\alpha_5)^2} k_{1\alpha} k_{3\beta} \bar{p}^2 .
\eqno(A.17)
$$
Substituting (A.16) and (A.17) into (A.12) 
will give an integral of the same form but with an opposite  
sign as that of (A.15) and so the expression in (A.12) is equal to
zero.  All the other diagrams in Fig. 12 can be handled in a similar fashion and
one will get zero contribution for each of them. 

Next we look at diagrams (9c)-(9f) and (9i)-(9l).  It is easy to see that after
being multiplied on by $k_{1\mu}$, (9d), (9e), (9j) and (9k) are reduced into
diagrams similar to that of Fig. 12.  For example, (9d) is reduced to (12a)
and (12b), similarly for (9e), (9j) and (9k).  Therefore, 
the diagrams (9d), (9e), (9j) and (9k) give zero contribution.
The only diagrams that would contribute to the anomaly amplitudes are
(9c), (9f), (9i) and (9l).  The anomaly amplitude from these diagrams are 
schematically shown in Fig. 13.  For 
example, (9c) is reduced into diagrams (13a) and (13b), etc.  Note that even
though Fig. 13 has the same topology as that of Fig. 12, the momentum
assignments are different.  For example, anomaly amplitude corresponding to
diagram (13a) is
$$
i \theta (1+\theta) f^2 \int \frac{d^4 p}{(2\pi)^4} \frac{d^4 q}{(2\pi)^2}
\frac{\bf N}{\bf D} ,
\eqno(A.18)
$$
where
$$
{\bf N} = (2 q + k_2)^\nu Tr \biggl[ \gamma_5 \gamma^\rho (\pslash + \qslash)
\pslash (\pslash + \qslash - \kslash_3) \biggr] ,
\eqno(A.19)
$$
$$
{\bf D} = [(q+k_2)^2 - 2\mu^2] q^2 [(p+q)^2 - m^2] [(p+q-k_3)^2 - m^2] 
(p^2 - m^2) .
\eqno(A.20)
$$
In the above, we can see that the momentum $q$ appears in four of the five
propagators in (A.20) as opposed to three in (A.3).  By restricting ourselves
to integrating over $q$ first and then over $p$ similar to the above, one gets 
a nonzero result for this diagrams which is equal to 
$\displaystyle \epsilon^{\nu\rho\sigma\sigma'}
\frac{k_{2\sigma} k_{3\sigma'}}{512 \pi^4} f^2 \theta (1+\theta)$.  Since each
diagram in Fig. 13 gives identical result, the total 
contribution from Fig. 13 is 
$$
\displaystyle
J_3 = \epsilon^{\nu\rho\sigma\sigma'} \frac{k_{2\sigma} k_{3\sigma'}}
{64 \pi^4} f^2 \theta (1+\theta) .
\eqno(A.21)
$$
 
\newpage

\begin{center}
\large\bf {Appendix B}
\end{center}

The renormalized vertex function $\Gamma_{V\psi\bar{\psi}}^{(r)}$ is equal
to the unrenormalized vertex function multiplied by a renormalization
constant $Z_{V\psi\bar{\psi}}$.  Customarily, one chooses the point of
subtraction at $\Delta = 0$, with both $p$ and $p'$ on the mass shell,
i.e. $p^2 = p^{'2} = m^2 $.  However, in the Abelian Higgs theory 
formulated in the Landau gauge, the $\phi_2$-field is massless.  As a 
consequence, the vertex function $\Gamma_{V\psi\bar{\psi}}$ is infrared
divergent when $p$ and $p'$ are both on the mass shell.  Take, for
example, diagram (6b).  The inverse of the two fermion propagators 
corresponding to this diagram have the following factors at $p_i^2 =
p_f^2 = m^2$, (where $p_f \equiv p'$ and $p_i \equiv p$)
$$\begin{array}{ll}
(p_i + q)^2 - m^2 &= 2 p_i \cdot q + q^2, \\
(p_f + q)^2 - m^2 &= 2 p_f \cdot q + q^2 .
\end{array}
$$
In addition, the inverse of the $\phi_2$-propagator is equal to $q^2$.  As
a consequence, the amplitude corresponding to diagram (6b) is divergent at
$q = 0$.

While $\Gamma_{V\psi\bar{\psi}}$ considered as a matrix is infrared 
divergent, some of its matrix elements are infrared finite.  In particular,
if we insert $\Gamma_{V\psi\bar{\psi}}$ into the two bispinors $\bar{u} 
(p_f)$ and $u (p_i)$, the amplitude is infrared finite.  As an example,
let us consider the amplitude corresponding to diagram (6b).  The numerator 
for the amplitude is 
$$
\displaystyle
\gamma_5 (\pslash_f + \qslash + m) \gamma^{\mu} \frac{1+2\theta+\gamma_5}{2}
(\pslash_i + \qslash + m) \gamma_5 .
$$
Since
$$\displaystyle
(\pslash_i + \qslash + m) \gamma_5 = \gamma_5 ( - \pslash_i - \qslash + m) ,
$$
this numerator operating on $u(p_i)$ becomes 
$$\displaystyle
\gamma_5 (\pslash_f + \qslash + m) \gamma^{\nu} \frac{1+2\theta+\gamma_5}{2} 
(-\qslash) ,
$$
which vanishes at $q = 0$.  Therefore, the matrix element of $\Gamma_{V\psi
\bar{\psi}}$ between the bispinors $\bar{u}(p_f)$ and $u(p_i)$ is infrared
finite.  As a matter of fact, we also have
$$
\bar{u}(p_f) \gamma_5 (\pslash_f + \qslash + m) = \bar{u} (p_f) (-\qslash)
\gamma_5 .
$$
Therefore, the numerator vanishes as $q^2$ when $q \rightarrow 0$.

Since the vertex function $\Gamma_{V\psi\bar{\psi}}$, considered as a matrix,
is infrared divergent, the point of subtraction will be chosen to be at
$\Delta = 0$ and $p^2 = p^{'2} = \Omega^2$, with $\Omega \ne m$.  The
renormalized vertex function $\Gamma_{V\psi\bar{\psi}}^{(r)}$ is therefore
equal to $Z_{V\psi\bar{\psi}} (\Omega^2, \Omega^2, 0)$ times the
unrenormalized vertex function.
 
\begin{center}
\bf
{(1) Diagram (6a)}
\end{center}

The diagram (6a) in Fig. 6 gives the following renormalized amplitude to
$\Gamma^\mu_{V\psi\bar{\psi}}$:

$$
  \Gamma^{\mu(a)}_{V\psi\bar{\psi}}(p_f, p_i)
  =
  -ig^2 \int \displaystyle {\frac {d^4q} {(2\pi)^4}}
\frac{n_a}{[(p_f+q)^2-m^2][(p_i+q)^2-m^2](q^2-M^2)}
$$
$$
- \mbox{term of subtraction},
\eqno{(B.1)}
$$
where
$$
  n_a
  \equiv
  \gamma^\sigma \frac{1+2\theta+\gamma_5}{2}(\pslash_f + \qslash+m)
  \gamma^\mu    \frac{1+2\theta+\gamma_5}{2}(\pslash_i + \qslash+m)
  \gamma^{\sigma'}\frac{1+2\theta+\gamma_5}{2}
  \left(
    g_{\sigma\sigma'}-\frac{q_\sigma q_{\sigma'}}{q^2}
  \right).
\eqno{(B.2)}
$$
By making use of (3.15)-(3.18), we may reduce the numerator $n_a$ in (B.2) to
$$\begin{array}{ll}
  n_a
  = &
\displaystyle  \biggl[ (1+\theta)^3 \frac{1-\gamma_5}{2} + \theta^3 
 \frac{1+\gamma_5}{2} \biggr] \biggl[-2 (\pslash_i + \qslash)
  \gamma^\mu(\pslash_f + \qslash)
  -
  \gamma^\mu \pslash_i \qslash
  -
  \qslash \pslash_f \gamma^\mu
  -
  \gamma_\mu q^2
  -
  \frac {\qslash \pslash_f \gamma^\mu \pslash_i \qslash}{q^2} \biggr] \\
& \displaystyle + \theta (1+\theta) m \biggl\{ \gamma^\sigma \gamma^\mu 
(\pslash_i + \qslash + m) \gamma^{\sigma'} \biggl[ (1+\theta)\frac{1+\gamma_5}
{2} + \theta \frac{1-\gamma_5}{2} \biggr] \\
 & {\hskip 3.0 cm} \displaystyle + \gamma^\sigma (\pslash_f + \qslash )
 \gamma^\mu \gamma^{\sigma'} \biggl[\theta \frac{1+\gamma_5}{2} + (1+\theta)
\frac{1-\gamma_5}{2} \biggr] \biggr\} \biggl( g_{\sigma\sigma'} - \frac{q_\sigma
 q_{\sigma'} }{q^2} \biggr) . \\
\end{array}
\eqno{(B.3)}
$$
We can further reduce (B.3) with the following identities
$$
\gamma_\sigma \gamma_\mu \gamma_\nu \gamma_\sigma = 4 g_{\mu\nu} 
\hskip 1 cm
\mbox{and} \hskip 1 cm
\gamma_\sigma \gamma_\mu \gamma_\sigma = - 2 \gamma_\mu .
$$
Substituting (B.3) into (B.1) and introducing Feynman parameters, we find that
the integration over $q$ is convergent.  We get
$$
\begin{array}{ll}
  \Gamma^{\mu(a)}_{V\psi\bar{\psi}}&
  =\displaystyle\frac{g^2}{16\pi^2}
  \left\{ \int d^3 \alpha \delta (1 - \sum \alpha)
  \frac{n_{a1}}{c_a}  + \int d^4\alpha \delta
  (1-\sum\alpha)
  \left(
    -
    \frac{n_{a2}}{d^2_a}
    +
    \frac{\pslash_i \gamma^\mu \pslash_f}{d_a}
  \right)\right\} \\
   & {\hskip 1.5 cm} \displaystyle \biggl[ (1+\theta)^3 \frac{1+\gamma_5}{2} + \theta^3 \frac{1-\gamma_5}{2} 
\biggr] \\
& \displaystyle + \frac{g^2}{16\pi^2} \theta(1+\theta) m \biggl\{  \int d^3 \alpha
\delta (1-\sum \alpha) \frac{n_{a3}}{c_a} - \int d^4 \alpha \delta (1-\sum\alpha)
\biggl( \frac{n_{a4}}{d_a^2} + \frac{n_{a5}}{d_a} \biggr) \biggr\} \\
&  -\mbox{term of subtraction}\\
\end{array}
\eqno{(B.4)}
$$
where
$$\begin{array}{ll}
n_{a1}  \equiv  & -2[(1-\alpha_1)\pslash_i-\alpha_2
\pslash_f]\gamma^\mu[(1-\alpha_2)\pslash_f-\alpha_1 \pslash_i] + \alpha_1
\pslash_i \pslash_f \gamma^\mu + \alpha_2 \gamma^\mu \pslash_i \pslash_f\\
&+\gamma^\mu(\alpha_1\alpha_3 p^2_i+\alpha_2\alpha_3
p_f^2+\alpha_1\alpha_2k^2),\\
\end{array} 
\eqno{(B.5)}
$$
$$
n_{a2} \equiv \alpha_1\alpha_2\gamma^\mu p^2_ip^2_f+\alpha^2_1 p^2_i
\pslash_i\pslash_f\gamma^\mu +\alpha^2_2 p_f^2 \gamma^\mu \pslash_i\pslash_f +
\alpha_1\alpha_2\pslash_i\pslash_f\gamma^\mu\pslash_i\pslash_f ,
\eqno{(B.6)}
$$
$$\begin{array}{ll}
n_{a3} \equiv & \displaystyle [4(1-\alpha_1) p_i^\mu - 4 \alpha_2 p_f^\mu
-2m\gamma^\mu + (\alpha_1 \pslash_i +\alpha_2 \pslash_f) \gamma^\mu] 
\biggl(\frac{1+2\theta+\gamma_5}{2} \biggr) \\
& + \displaystyle [4 (1-\alpha_2) p_f^\mu - 4 \alpha_1 p_i^\mu + \gamma^\mu
(\alpha_1 \pslash_i + \alpha_2 \pslash_f)]\biggl(\frac{1+2\theta-\gamma_5}{2}
\biggr) ,
\end{array} 
\eqno(B.7)
$$
$$\begin{array}{ll}
n_{a4} \equiv & \displaystyle (\alpha_1\pslash_i +\alpha_2\pslash_f) \pslash_f
\gamma^\mu (\alpha_1\pslash_i + \alpha_2 \pslash_f)
\biggl(\frac{1+2\theta-\gamma_5}{2} \biggr) \\
& + \displaystyle[m(\alpha_1\pslash_i +\alpha_2 \pslash_f) \gamma^\mu (\alpha_1
+ \alpha_2 \pslash_f) + (\alpha_1 \pslash_i + \alpha_2 \pslash_f) \gamma^\mu 
\pslash_i (\alpha_1 \pslash_i + \alpha_2 \pslash_f)]
\biggl( \frac{1+2\theta+\gamma_5}{2} \biggr) , 
\end{array}
\eqno(B.8)
$$
$$
n_{a5} \equiv  \displaystyle (2 p_i^\mu - m \gamma^\mu) 
\biggl(\frac{1+2\theta+\gamma_5}{2} \biggr) 
 + 2 p_f^\mu \biggl(\frac{1+2\theta-\gamma_5}{2} \biggr) , 
\eqno(B.9)
$$
$$
c_{a}  \equiv  \alpha_1\alpha_3 p^2_i + \alpha_2\alpha_3 p_f^2 +
\alpha_1\alpha_2 k^2 - (\alpha_1+\alpha_2)m^2-\alpha_3 M^2,
\eqno{(B.10)}
$$
$$
d_{a}  \equiv  \alpha_1(\alpha_3+\alpha_4)p^2_i +
\alpha_2(\alpha_3+\alpha_4)p_f^2 + \alpha_1\alpha_2
k^2-(\alpha_1+\alpha_2)m^2-\alpha_3 M^2.
\eqno{(B.11)}
$$

To obtain the term of subtraction in (B.4), we set $\displaystyle p_i=p_f=P_0$
with $\displaystyle P^2_0=\Omega^2$
in the integrals of (B.4).  We get, at the point of substraction, 
$$
c_a = (\alpha_1 + \alpha_2)\alpha_3 \Omega^2 
-(\alpha_1+\alpha_2) m^2-\alpha_3 M^2,
\eqno{(B.12)}
$$
$$
d_a = (\alpha_1+\alpha_2)(\alpha_3+\alpha_4) \Omega^2 - (\alpha_1+\alpha_2) m^2
- \alpha_3 M^2 ,
\eqno(B.13)
$$
$$
n_{a1} = -4\alpha^2_3 P_0^\mu \Pslash_0  + (1+\alpha^2_3)\Omega^2\gamma^\mu,
\eqno{(B.14)}
$$
$$
n_{a2} = (\alpha_1+\alpha_2)^2 \Omega^4\gamma^\mu ,
\eqno{(B.15)}
$$
$$\begin{array}{ll}
n_{a3} = & \displaystyle [4\alpha_3 P_0^\mu- 2m\gamma^\mu + (\alpha_1+\alpha_2) \Pslash_0 \gamma^\mu
] \biggl(\frac{1+2\theta+\gamma_5}{2} \biggr) \\
& \displaystyle + [4\alpha_3
P_0^\mu + \gamma^\mu (\alpha_1+\alpha_2) \Pslash_0] \biggl(\frac{1+2\theta-\gamma_5}{2} \biggr)  , 
\end{array}
\eqno(B.16)
$$
$$
\begin{array}{ll}
n_{a4} =& \displaystyle (\alpha_1+\alpha_2)^2 \biggl\{ (\Omega^2 \Pslash_0 \gamma^\mu + 2m P^\mu_0
\Pslash_0 -m\Omega^2 \gamma^\mu )\biggl(\frac{1+2\theta+\gamma_5}{2} \biggr) \\ 
& \displaystyle + \Omega^2 \gamma^\mu \Pslash_0  
\biggl( \frac{1+2\theta-\gamma_5}{2} \biggr) \biggr\} ,\\ 
\end{array}
\eqno(B.17)
$$
$$
\displaystyle
n_{a5} = 2 (1+2\theta) P_0^\mu - m \gamma^\mu \biggl( \frac{1+2\theta+\gamma_5}
{2} \biggr) .
\eqno(B.18)
$$
and
$$
\Pslash_0 \gamma^\mu \Pslash_0 = 2P^\mu_0 \Pslash_0 - \Omega^2 \gamma^\mu .
\eqno{(B.19)}
$$

The term of subtraction in (B.4) 
is proportional to $\displaystyle \gamma^\mu \biggl(\frac{1+2\theta+\gamma_5}{2}
\biggr)$ with the constant of proportionality equal to such a value that 
$\alpha_+$ defined in (5.26) is equal to zero at $p_i = p_f = P_0$.  Thus, we 
find, using (B.12)-(B.18), that the term of subtraction in (B.4) is, 
$$\begin{array}{ll}
&\displaystyle
\frac{g^2 \gamma^\mu (1+2\theta+\gamma_5)}{32\pi^2}  \biggl\{ 
\displaystyle \int d^3 \alpha \delta (1-\sum \alpha)
\frac{ \displaystyle \frac{(1+2\theta+2\theta^2)}{2} (1+\alpha_3^2) \Omega^2 - \theta (1+
\theta)(1+\alpha_1+\alpha_2) m^2}{(\alpha_1+\alpha_2)\alpha_3\Omega^2 - (\alpha_1
+\alpha_2)m^2 - \alpha_3 M^2} \\
& \displaystyle
- \int d^4 \alpha \delta(1 - \sum \alpha) \frac{\displaystyle \frac{(1+2\theta+2\theta^2)}{2}
[(\alpha_1+\alpha_2)
(\Omega^2- m^2) - \alpha_3 M^2] \Omega^2 - \theta(1+\theta) m^2 \Omega^2}{[(\alpha_1+\alpha_2)
(\alpha_3+\alpha_4)\Omega^2 - (\alpha_1+\alpha_2)m^2 - \alpha_3 M^2]^2}
\biggr\} \\
\end{array}
\eqno(B.20)
$$

Next we derive the asymptotic form of $\displaystyle\Gamma^{\mu(a)}_{V\psi\bar{\psi}}$ in
the limit $p\rightarrow\infty$ with $k$ fixed.  We shall only keep the part of
this asymptotic form which is relevant to the anomaly amplitude.  The terms 
that are proportional to $m$, i.e. lines 2 and 3 in (B.4) do not therefore,
contribute to the anomaly amplitude.  Let us first
observe that in $J^{2}$ as given by (4.5), $\displaystyle
\Gamma^{\mu}_{V\psi\bar{\psi}}\displaystyle(p_f,p_i)$
is always inserted between two propagators.  Take, for example, the first
trace in (4.5b).  Let us insert for $\Gamma^\mu$ in this trace a next-lowest
order term.  Then we need to keep only the lowest-order terms for $S$ and
$\Gamma^\rho$.  We shall drop $m$ in $S$, as the limit $p\rightarrow\infty$
will be eventually taken and one has only to worry about the linearly divergent
piece.  Thus the first trace in (4.5b) is approximately
$$
(\frac{i}{p^2})^2 Tr \biggl[\Gamma^\nu(p+k_3, p-k_1)\frac{(1+2\theta+\gamma_5)}{2}
\pslash
\gamma^\rho \frac{(1+2\theta+\gamma_5)}{2}(\pslash+k_3)\biggr].
\eqno{(B.21)}
$$
We shall keep only the terms in the trace which are linearly proportional to
$\gamma_5$, as other terms cancel by an extension of Furry's theorem.  Since
the anomaly amplitude is finite, we anticipate the trace in (B.21) be
asymptotically proportional to $p$.  (Individual terms may actually have
factors of $\ln p^2$, but such factors cancel one another.)  By counting
dimensions, we see that terms in the trace in (B.21) must be, asymptotically,
proportional to one factor of k, where $k$ is an external momentum, in
addition to a factor of $p$.  Let us now
make approximations for the numerators $n_{a1}$ and $n_{a2}$ by first moving
$\pslash_i$ to the right and $\pslash_f$ to the left.  For example, for
$n_{a1}$ in (B.5), a term $\displaystyle\pslash_i\gamma^\mu
\pslash_f$ will be expressed as
$$
2p_i^\mu \pslash_f - 2(p_i\cdot p_f)\gamma^\mu + 2 p^\mu_f \pslash_i - \pslash_f
\gamma^\mu \pslash_i.
\eqno{(B.22)}
$$
Let us imagine inserting (B.22) into (B.21), with $\mu, p_i$ and $p_f$
identified with $\nu, p-k_1$ and $p+k_3$, respectively.  The third term in
(B.22) gives a term proportion to:
$$
p^\mu_f Tr {\biggl [} \gamma_5\gamma^\rho
\pslash(\pslash+\kslash_3)(\pslash-\kslash_1){\biggr ]} = -p^\mu_f Tr
{\biggl [}\gamma_5\gamma^\rho \pslash\kslash_3\kslash_1{\biggr ]} ,
$$
which is proportional to two powers of k.  This means that a power of $p$ has
been lost (we remind the reader that there is a factor $1/p^2$ coming from the
denominator $c_a$.)  We shall therefore drop the third term in (B.22), which
does not contain a factor $\gamma^\mu$.  We
also drop the first term in (B.22) for the same reason.  Thus we get 
$$
\pslash_i\gamma^\mu \pslash_f \approx -2p^2\gamma^\mu -
\pslash_f\gamma^\mu \pslash_i.
\eqno{(B.23)}
$$
Similarly,
$$
\pslash_f\gamma^\mu \pslash_f \approx -p^2\gamma^\mu
\eqno{(B.24)}
$$
and
$$
\pslash_i\gamma^\mu \pslash_i \approx -p^2\gamma^\mu
\eqno{(B.25)}
$$
$$
\pslash_i \pslash_f\gamma^\mu \approx 2p^2\gamma^\mu + \pslash_f \gamma^\mu
\pslash_i ,
\eqno{(B.26)}
$$
$$
\gamma^\mu \pslash_i \pslash_f \approx 2p^2\gamma^\mu + \pslash_f\gamma^\mu
\pslash_i.
\eqno{(B.27)}
$$
Consequently
$$
n_{a1} \approx \gamma^\mu(2+\alpha_3+\alpha^2_3)p^2 + \pslash_f \gamma^\mu
\pslash_i(1+\alpha_3).
\eqno{(B.28)}
$$
Similarly,
$$
n_{a2} \approx 2\gamma^\mu(p^2)^2(\alpha_1+\alpha_2)^2 + \pslash_f \gamma^\mu
\pslash_i p^2(\alpha_1+\alpha_2)^2 .
\eqno{(B.29)}
$$

One must resist the temptation of dropping the last three terms in (B.10) and
those in 
(B.11),  as the resulting integrals in (B.4) become divergent.  This divergence
means that the first integral in (B.4) has a $\ln p^2$ term, as the divergence
at $\alpha_3=0$ is logarithmic.  Similar considerations apply for the second
integral in (B.4).  It is legitimate, however, to drop the term $\alpha_3 M^2$
in (B.10) and (B.11), and replace $p^2_i$ and $p^2_f$ by $p^2$.  We obtain
$$
c_a \approx (\alpha_1+\alpha_2)\alpha_3 p^2 + \alpha_1\alpha_2 k^2
-(\alpha_1+\alpha_2)m^2
\eqno{(B.30)}
$$
and
$$
d_a \approx (\alpha_1+\alpha_2)(\alpha_3+\alpha_4)p^2+\alpha_1\alpha_2 k^2
-(\alpha_1+\alpha_2)m^2.
\eqno{(B.31)}
$$
This approximation is allowed because $\alpha_3 M^2$ is by an order of $p^2$ smaller
than $(\alpha_1+\alpha_2)\alpha_3 p^2$, as $(\alpha_1+\alpha_2)$ is the order
of unity in the region of integration which contributes to the asymptotic
form.

Next we make the following change of variables for the second integral in
(B.4):
$$\begin{array}{ll}
\alpha_3+\alpha_4  & \equiv \rho,\\
\alpha_3 & \equiv \rho x,\\
\alpha_4 & \equiv \rho(1-x).\\
\end{array}
\eqno{(B.32)}$$
It follows from (B.32) that
$$
d\alpha_3 d\alpha_4 = \rho d \rho dx.
$$
With the approximation (B.31), the integrands of the second integral in (B.4)
is independent of $x$.  Thus the intgration over $x$ simply yields unity.  We
shall rename $\rho$ as $\alpha_3$, and obtain (B.4) as
$$\begin{array}{ll}
\Gamma^{\mu(a)}_{V\psi\bar{\psi}}(p_f,p_1) \approx \displaystyle\frac{g^2}{16\pi^2}\int & \displaystyle 
d^3\alpha\delta(1-\sum\alpha) \biggl[ \frac{p^2\gamma^\mu(2-\alpha_3+\alpha^2_3)
+\pslash_f \gamma^\mu \pslash_i}{(1-\alpha_3)\alpha_3p^2+\alpha_1\alpha_2 k^2 
- (\alpha_1+\alpha_2)m^2} \\
&\displaystyle - \frac{(2p^2 \gamma^\mu + \pslash_f \gamma^\mu \pslash_i)p^2
\alpha_3(1-\alpha_3)^2} {[(1-\alpha_3)\alpha_3p^2+\alpha_1\alpha_2 k^2 -
(\alpha_1+\alpha_2)m^2]^2} \biggr]
\biggl(\frac{1+2\theta+\gamma_5}{2} \biggr)^3 \\ 
& -\mbox{term of
subtraction}. 
\end{array}
\eqno{(B.33)}
$$
Eq. (B.33) can be rewritten as
$$\begin{array}{ll}
\Gamma^{\mu(a)}_{V\psi\bar{\psi}}(p_f,p_i)& \displaystyle \approx \frac{g^2}{16\pi^2} \int d^3
\alpha \delta(1-\sum\alpha) \biggl[ \frac{p^2\gamma^\mu\alpha_3(1+\alpha_3)+\alpha_3 \pslash_f \gamma^\mu
\pslash_i} {(1-\alpha_3)\alpha_3 p^2 + \alpha_1 \alpha_2 k^2 -
(\alpha_1+\alpha_2)m^2} \\
& \displaystyle + \frac{[\alpha_1\alpha_2 k^2 -
(\alpha_1+\alpha_2)m^2](1-\alpha_3)(2p^2\gamma^\mu + \pslash_f \gamma^\mu
\pslash_i)}{[(1-\alpha_3)\alpha_3 p^2 + \alpha_1\alpha_2 k^2 -
(\alpha_1+\alpha_2)m^2]^2} \biggr] \biggl(\frac{1+2\theta+\gamma_5}{2} 
\biggr)^3 \\
& -\mbox{term of subtraction.}
\end{array}
\eqno{(B.34)}
$$
We can now drop $\alpha_1\alpha_2 k^2 - (\alpha_1+ \alpha_2)m^2$ in the
denominator of the first term of (B.34), the resulting integral being convergent.
 This cannot be done, however, for the second term of (B.34), as the resulting
integral remains linearly divergent at $\alpha_3 = 0$.  Instead, this term is
asymptotically given by the contribution at $\alpha_3=0$, and we may
approximate $1-\alpha_3$ by $1$ and carry out the integration.  We get
$$\begin{array}{ll}
\Gamma^{\mu(a)}_{V\psi\bar{\psi}}(p_f,p_i) & \approx
\displaystyle{\frac{g^2}{16\pi^2}\biggl[\int d^3 \alpha \delta(1-\sum\alpha) \displaystyle\frac{\gamma^\mu(1+\alpha_3) + \pslash_f
\gamma^\mu \pslash_i/p^2}{1-\alpha_3} }\\
&{\hskip 1.5cm}\displaystyle {+ 2\gamma^\mu + \pslash_f \gamma^\mu \pslash_i/p^2
\biggr]
\biggl(\frac{1+2\theta+\gamma_5}{2}\biggr)^3
-\mbox{term of subtraction} } \\
& = \displaystyle\frac{g^2}{16\pi^2}(\frac{7}{2} \gamma^\mu + 2 \pslash_f \gamma^\mu
\pslash_i/p^2) \biggl(\frac{1+2\theta+\gamma_5}{2}\biggr)^3 - \mbox{term of
subtraction} ,\\
\end{array}
\eqno{(B.35)}
$$
which is the part of the asymptotic form of
$\Gamma^{\mu(a)}_{V\psi\bar{\psi}}$ contributing to the anolamy amplitude.

The term of subtraction in (B.35) is 
the coefficient of the term of subtraction (B.20) with
$\alpha_3$ replaced by $(1-\alpha)$, which depends on the masses $m$ and $M$.  The
first term in (B.35) comes from the first term in (B.1) which, as we have
mentioned, is a convergent integral, and has no mass dependences.  There is no
$\ln p^2$ term, as terms of this order cancel.

\begin{center}
\bf {(2) Diagram (6b) }
\end{center}

Diagram (6b) gives the following renormalized amplitude to
$\Gamma^\mu_{V\psi\bar{\psi}}$ :
$$
-if^2 \int \frac{d^4q}{(2\pi)^4} \frac{1}{q^2} \gamma_5 \frac{\pslash_f +
\qslash+m}{(p_f+q)^2-m^2} \gamma^\mu \biggl(\frac{1+2\theta+\gamma_5}{2}\biggr)
\frac{\pslash_i+ \qslash+m}{(p_i+q)^2-m^2}\gamma_5
-\mbox{term of subtraction.}
\eqno{(B.36a)}
$$
One has to be careful about the term of subtraction here.  If we want to 
make a subtraction according to the choice $\bar{\alpha}_+ = 1$, then the
divergent part that is proportional to $\displaystyle \gamma^\mu \frac{1-
\gamma_5}{2}$ is only partially cancelled by the term of subtraction introduced
here.  The remaining divergence will be cancelled by the divergence of the same
form in amplitudes corresponding to diagrams (6f) and (6g).  The term of 
subtraction, as we introduce here, takes the following form 
$$\begin{array}{ll}
& - \displaystyle 
\frac{if^2\theta}{(1+\theta)} \int d^3 \alpha \delta (1 - \sum \alpha) \int 
\frac{d^4 q}{(2\pi)^4} \frac{ q^2 + 2\alpha_3^2\Omega^2 }
{[q^2 + (\alpha_1+\alpha_2) \alpha_3 \Omega^2 - (\alpha_1+\alpha_2) m^2 ]^3} 
\gamma^\mu \biggl( \frac{1+2\theta+\gamma_5}{2} \biggr) \\
& - \displaystyle \frac{f^2}{16\pi^2} \int d^3 \alpha \delta (1-\sum\alpha)
\frac{m^2}{(\alpha_1+\alpha_2)\alpha_3 \Omega^2 - (\alpha_1+\alpha_2) m^2} 
\gamma^\mu \biggl( \frac{1+2\theta+\gamma_5}{2} \biggr) .
\end{array}
\eqno(B.36b)
$$
In the asymptotic limit, i.e. $p^2 \rightarrow \infty$, we may drop $m$ in the
numerator of the integral in (B.36a).  One then introduces Feynman parameters as
usual and make a shift in q, i.e. $q \rightarrow \bar{q} - \alpha_1 p_i - \alpha_2 
p_f$.   The first term in (B.36a) that is relevant to the anomaly amplitude then reads
$$
\begin{array}{ll}
 -i f^2 \displaystyle \int d^3 \alpha \delta(1-\sum \alpha) \int \frac{d^4 \bar{q}}
{(2\pi)^4} & \displaystyle \frac{2[(1-\alpha_2) \pslash_f - \alpha_1 \pslash_i]\gamma^\mu
[(1-\alpha_1)\pslash_i - \alpha_2 \pslash_f] - \bar{q}^2 \gamma^\mu} { [\bar{q}^2 +
\alpha_1 \alpha_3 p_i^2 + \alpha_2 \alpha_3 p_f^2 + \alpha_1 \alpha_2 k^2
- (\alpha_1 + \alpha_2) m^2 ]^3 } \\
& \displaystyle \biggl( \frac{1+2\theta-\gamma_5}{2} 
\biggr) . 
\end{array}
\eqno(B.37)
$$
From the arguments following (B.22), it is easy to see that only the terms 
$\bar{\qslash} \gamma^\mu \bar{\qslash}$ and $\pslash_f \gamma^\mu \pslash_i$ in (B.36)
contribute to the anomaly amplitude.
Using (B.23)-(B.27), the first term in the numerator of (B.37) gives
$$
- f^2 \biggl( \frac{\gamma^\mu}{2} + \frac{\pslash_f \gamma^\mu \pslash_i}
{p^2} \biggr) \biggl(\frac{1+2\theta-\gamma_5}{2}\biggr) ,
\eqno(B.38)
$$
while the second term in the numerator of (B.37), together with the first
line in (B.36b) and part of the subtraction term in (B.45) 
as will be discussed below in section (5) gives
$$
\begin{array}{ll}
\displaystyle - \frac{f^2}{16\pi^2}  \int d^3 \alpha \delta (1-\sum \alpha)  
& \biggl\{ \displaystyle \ln \biggl( \frac{\alpha_3 p^2}{\alpha_3 \Omega^2 - m^2} \biggr) 
\gamma^\mu \biggl( \frac{1+2\theta-\gamma_5}{2} \biggr) \\
& \displaystyle +\biggl( \frac{1+2\theta}{1+\theta}\biggr) \ln \biggl[
\frac{(\alpha_1+\alpha_2) \alpha_3 \Omega^2 - (\alpha_1+\alpha_2) m^2}
{(\alpha_1+\alpha_2) \alpha_3 \Omega^2 - 2\alpha_1 \mu^2 - \alpha_3 m^2}
\biggr] \gamma^\mu \biggl( \frac{1-\gamma_5}{2} \biggr) \\ 
& \displaystyle + \biggl( \frac{1}{1+\theta} \biggr) \frac{\theta \alpha_3^2 \Omega^2 - (1+\theta) m^2}{(\alpha_1+\alpha_2) 
\alpha_3 \Omega^2 - (\alpha_1+\alpha_2) m^2} \gamma^\mu \biggl(
\frac{1+2\theta+\gamma_5}{2} \biggr)\biggr\} .
\end{array}
\eqno(B.39)
$$
The contribution of diagram (6b) to $\Gamma_{V\psi\bar{\psi}}^\mu$ relevant
to the anomaly amplitude is therefore equal to 
$$
\begin{array}{ll}
\,\,\,\,\,\,-  \displaystyle f^2 \biggl(\frac{\gamma^\mu}{2} + \frac{\pslash_f \gamma^\mu \pslash_i}
{p^2} \biggr) \,\,\frac{1}{2} \,\,( &1 + 2 \theta -\gamma_5\,\,) \\ 
\,\,-  \displaystyle \frac{f^2}{16\pi^2} \int d^3 \alpha \delta (1-  \sum \alpha)
& \biggl\{ \displaystyle \ln \biggl(  \frac{\alpha_3 p^2}{\alpha_3 \Omega^2 -
m^2} \biggr)
\gamma^\mu \biggl( \frac{1+2\theta-\gamma_5}{2} \biggr) \\
& \displaystyle +\biggl( \frac{1+2\theta}{1+\theta}\biggr) \ln \biggl[
\frac{(\alpha_1+\alpha_2) \alpha_3 \Omega^2 - (\alpha_1+\alpha_2) m^2}
{(\alpha_1+\alpha_2) \alpha_3 \Omega^2 - 2\alpha_1 \mu^2 - \alpha_3 m^2}
\biggr] \gamma^\mu \biggl( \frac{1-\gamma_5}{2} \biggr) \\
& \displaystyle + \biggl( \frac{1}{1+\theta} \biggr) \frac{\theta \alpha_3^2
\Omega^2 - (1+\theta) m^2}{(\alpha_1+\alpha_2)
\alpha_3 \Omega^2 - (\alpha_1+\alpha_2) m^2} \gamma^\mu \biggl(
\frac{1+2\theta+\gamma_5}{2} \biggr) \biggr\} .
\end{array}
\eqno(B.40)
$$

\begin{center}
\bf {(3) Diagram (6c)}
\end{center}

Diagram (6c) gives the following renormalized amplitude to
$\Gamma^\mu_{V\psi\bar{\psi}}$:
$$
if^2 \int \frac{d^4q}{(2\pi)^4} \frac{1}{q^2 - 2\mu^2}
\frac{\pslash_f+\qslash+m}{(p_f+q)^2-m^2} \gamma^\mu 
\biggl(\frac{1+2\theta+\gamma_5}{2}\biggr)
\frac{\pslash_i+\qslash+m}{(p_i+q)^2-m^2}
-\mbox{term of subtraction.}
\eqno{(B.41a)}
$$
Analogous to the discussion for (6b), the term of subtraction in (B.41a) is found to be
$$\begin{array}{ll}
&  - \displaystyle 
\frac{i f^2\theta}{(1+\theta)} \int d^3 \alpha \delta(1-\sum\alpha) \int \frac{d^4 q}
{(2\pi)^4} \frac{(q^2+2\alpha_3^2\Omega^2) \gamma^\mu \displaystyle ( \frac{1+2\theta+\gamma_5}{2})}{[q^2 + (\alpha_1+\alpha_2)\alpha_3 \Omega^2 - (\alpha_1
+\alpha_2) m^2 - 2\alpha_3 \mu^2]^3}  \\
& - \displaystyle \frac{f^2}{16\pi^2} \int d^3 \alpha \delta (1-\sum\alpha) 
\frac{m^2}{ (\alpha_1+\alpha_2) \alpha_3 \Omega^2 - (\alpha_1+\alpha_2) m^2
- 2 \alpha_3 \mu^2} \gamma^\mu \biggl( \frac{1+2\theta+\gamma_5}{2} \biggr)  .
\end{array}
\eqno(B.41b)
$$
In a similar fashion, the contribution of diagram (6c) is found to be
$$
\begin{array}{ll}
- & \displaystyle   f^2 \biggl(\frac{\gamma^\mu}{2} + \frac{\pslash_f 
\gamma^\mu \pslash_i}{p^2} \biggr) \biggl(\frac{1+2\theta-\gamma_5}{2} \biggr)
\\
- & \displaystyle \frac{f^2}{16\pi^2} \int d^3 \alpha \delta (1-\sum \alpha) 
\biggl\{ \ln \biggl[ \frac{ (\alpha_1+\alpha_2) \alpha_3 p^2}{ (\alpha_1 +
\alpha_2) \alpha_3 \Omega^2 - (\alpha_1+\alpha_2) m^2 - 2 \alpha_3 \mu^2}
\biggr] \gamma^\mu \biggl(\frac{1+2\theta-\gamma_5}{2} \biggr) \\
& +\displaystyle \biggl( \frac{1+2\theta}{1+\theta} \biggr) \ln \frac{ (\alpha_1
+\alpha_2) \alpha_3 \Omega^2 - (\alpha_1+\alpha_2) m^2 - 2\alpha_3 \mu^2}
{(\alpha_1+\alpha_2) \alpha_3 \Omega^2 - 2\alpha_1 \mu^2 - \alpha_3 m^2}
\gamma^\mu \biggl( \frac{1-\gamma_5}{2} \biggr) \\
& + \displaystyle \biggl( \frac{1}{1+\theta}\biggr) \frac{\theta \alpha_3^2
\Omega^2 - (1+\theta) m^2}{ (\alpha_1+\alpha_2) \alpha_3 \Omega^2 - (\alpha_1
+\alpha_2) m^2 - 2\alpha_3 \mu^2} \gamma^\mu \biggl( \frac{1+2\theta+\gamma_5}
{2} \biggr) \biggr\}  .
\end{array}
\eqno(B.42)
$$

\begin{center}
\bf {(4)  Diagrams (6d) and (6e)}
\end{center}

Diagram (6d) and diagram (6e) give the following amplitude to
$\Gamma^\mu_{V\psi\bar{\psi}}$ valid to the next-lowest order:

$$\begin{array}{ll}
\displaystyle 2ig^2 M \int
\displaystyle\frac{d^4q}{(2\pi)^4} \biggl[\displaystyle\frac{\pslash_i+\qslash+m}{(p_i+q)^2-m^2}\gamma^\nu\displaystyle
\frac{(1+2\theta+\gamma_5)}{2} + \gamma^\nu\displaystyle\frac{(1+2\theta+\gamma_5)}{2}
\displaystyle\frac{\pslash_f-\qslash+m}{(p_f-q)^2-m^2}\biggr]&\\
\hspace{.50in}\displaystyle\frac{g^{\mu\nu}-q^\mu
q^\nu/q^2}{q^2-M^2}\displaystyle\frac{1}{(k+q)^2-2\mu^2}-\mbox{term of subtraction.}&\\
\end{array}
\eqno{(B.43a)}
$$
The term of subtraction in (B.43a) is evaluated to be
$$\begin{array}{ll}
\displaystyle 
\frac{-2g^2 M m}{16\pi^2} \biggl{\{} & \displaystyle
\int d^3 \alpha \frac{ 2 \, \delta (1-\sum\alpha)}{\alpha_1 m^2 - \alpha_1
(1-\alpha_1) \Omega^2 + 2 \alpha_2 \mu^2 + \alpha_3 M^2} \\ 
& \displaystyle
+ \int d^4 \alpha \frac{\delta (1-\sum\alpha)}{\alpha_1 m^2 - \alpha_1 (1-\alpha_1)
\Omega^2 + 2\alpha_2\mu^2 + \alpha_3 M^2} \biggr{\}}
\gamma^{\mu} \biggl( \frac{1+2\theta+\gamma_5}{2} \biggr) .
\end{array}
\eqno(B.43b)
$$
Because of the factor of $M$ outside the integral sign, we conclude by counting
dimension that the first term in (B.43a) does not contribute to the anomaly 
amplitude.  Thus the expression in (B.43a) is replaced by the negative of
(B.43b).

\begin{center}
\bf {(5) Diagrams (6f) and (6g) }
\end{center}

Diagrams (6f) and (6g) give the following renormalized amplitude to
$\displaystyle\Gamma^{\mu}_{V\psi\bar{\psi}}$:
$$\begin{array}{ll}
\displaystyle-if^2 \int
\displaystyle\frac{d^4q}{(2\pi)^4}(2q+k)^\mu \biggl[\displaystyle\frac{\pslash_i+\qslash-m}{(p_i+q)^2-m^2}\displaystyle\frac{1}{q^2-2\mu^2}\displaystyle\frac{1}{(k+q)^2}&\\
\hspace{.50in}+\displaystyle \frac{\pslash_i+\qslash+m}{(p_i+q)^2-m^2} \displaystyle\frac{1}{q^2}
\displaystyle\frac{1}{(k+q)^2-2\mu^2} \biggr]\gamma_5 - \mbox{term of subtraction.}&\\
\end{array}
\eqno{(B.44)}
$$
where, in the first term of the integrand, there is a matrix $\gamma_5$ which
we have moved to the rightest position.  The integral in (B.44) has a
logarithmic ultraviolet divergence, which is not entirely cancelled by the
term of subtraction in (B.44).  This is because the former is proportional to 
$\displaystyle \gamma^\mu \gamma_5$,  which can be written as:
$$
\gamma^\mu \frac{1}{2}(1+\gamma_5)-\gamma^\mu \frac{1}{2}(1-\gamma_5).
$$
The term of subtraction that we introduce here only cancels the divergent
part of the form $\displaystyle \gamma^\mu \biggl(\frac{1+\gamma_5}{2}\biggr)$.
The uncancelled divergence, in the form of $\displaystyle \gamma^\mu
\frac{1}{2}(1-\gamma_5)$, is cancelled by the divergence of the same form in
amplitudes corresponding to diagrams (6b) and (6c) that still remains as we have discussed above.  An amplitude of such a
form is ultraviolet finite and requires no subtraction, as we have concluded in Sec. 2.

Let us single out the terms in (B.44) which contribute to the anomaly
amplitude.  First of all, we drop $m$ in the numerators in (B.44). 
Then the relevant part in (B.44) is
$$\begin{array}{ll}
\displaystyle -if^2 \int & \displaystyle\frac{d^4q}{(2\pi)^4}(2q+k)^\mu
\displaystyle\frac{\pslash_i+\qslash}{(p_i+q)^2-m^2} \biggl[\displaystyle\frac{1}{q^2-2\mu^2}\displaystyle\frac{1}{(k+q)^2}+ \displaystyle\frac{1}{q^2}
\displaystyle\frac{1}{(k+q)^2-2\mu^2}\biggr] \gamma_5 \\
& -\mbox{term of subtraction.}\\
\end{array}
\eqno{(B.45)}
$$
Introducing Feynman parameters and calling
$$\begin{array}{ll}
q &\equiv \bar{q}+\delta q, \\
\delta q &\equiv -\alpha_3p_i-\alpha_2 k,
\end{array}
$$
we find that, since an amplitude of the form of $\displaystyle p^\mu\pslash
\frac{1}{2}(1 \pm \gamma_5)$ does not contribute to the anomaly amplitude, as we
have discussed in the paragraph following (B.22), we may drop all terms in
(B.45) involving $\delta q$.  The subtraction term is given by
$$
- \displaystyle \frac{if^2}{(1+\theta)} \int d^3 \alpha \delta (1-\sum \alpha)  
\int \frac{d^4 q}{(2\pi)^4} \frac{ 2q^2}{[q^2 + (\alpha_1+\alpha_2) \alpha_3
\Omega^2 - 2 \alpha_1 \mu^2 - \alpha_3 m^2 ]^3} \gamma^\mu \biggl(
\frac{1+2\theta+\gamma_5}{2} \biggr) \,\, .
\eqno(B.46)
$$
We break up the last factor in (B.46) into two terms: 
$$
\frac{1+2\theta+\gamma_5}{2} = (1+\theta) \gamma_5 + (1+2\theta) \biggl(
\frac{1-\gamma_5}{2} \biggr) .
\eqno(B.47)
$$
Then the subtraction term in (B.46) becomes a sum of two terms.  The first term 
combines with (B.45) to  
give the following contribution to the anomaly amplitude :
$$
 \displaystyle\frac{2f^2}{16\pi^2} \int d^3\alpha \delta(1-\sum\alpha)
\ln {\biggl [} \frac{\alpha_3 m^2 - (\alpha_1 + \alpha_2) \alpha_3 \Omega^2
+2\alpha_1\mu^2}{-(\alpha_1+\alpha_2)\alpha_3 p^2} {\biggr ]} \gamma^\mu
 \gamma_5  . 
\eqno{(B.48)}
$$
The remaining term  
combines with the subtraction terms in (B.36b) of diagram (6b) and in (B.41b)
of diagram (6c) to give a finite
contribution to the anomaly amplitude as given in (B.40) and (B.42) above.

\newpage
\begin{center}
\large \bf {Appendix C}
\end{center}

In this appendix, we discuss the lowest-order radiative corrections of the
fermion propagator, with special emphasis on their contributions to the
anomaly amplitude.  The relevant diagrams are illustrated in Fig. 11.

\begin{center}
\bf {(1)  Diagram (11a) }
\end{center}

The renormalized amplitude corresponding to diagram 11a is:
$$
{\scriptstyle\sum_a^{(r)}}(p) = if^2\int \frac{d^4q}
{(2\pi)^4}\frac{\pslash-\qslash+m}{(p-q)^2-m^2}
\frac{1}{q^2-2\mu^2}-\mbox{term of subtraction.}
\eqno{(C.1
)}$$
Introducing Feynman parameters, we get
$$
{\scriptstyle\sum^{(r)}_a}(p) = if^2\int^1_0 d\alpha\int \displaystyle\frac{d^4\bar{q}}{(2\pi)^4}\displaystyle\frac{(1-\alpha)\pslash+m}{[\bar{q}^2+\alpha(1-\alpha)p^2-\alpha
m^2-2(1-\alpha)\mu^2]^2}-\mbox{term of subtraction,}
\eqno{(C.2)}
$$
where
$$
\bar{q}=q-\alpha p.
$$
The integral in (C.2) is logarithmically divergent.  After we make a
subtraction it is equal to:
$$
\pslash A_a(p^2)+ m B_a(p^2),
\eqno{(C.3)}
$$
where
$$
A_a(p^2) = \displaystyle\frac{-f^2}{(16\pi^2)} \int^1_0 d\alpha(1-\alpha) 
\, \ln \biggl[ \frac{\alpha^2
m^2+2(1-\alpha)\mu^2}{-\alpha(1-\alpha)p^2+\alpha m^2+2(1-\alpha)\mu^2} \biggr],
\eqno{(C.4)}
$$
and
$$
B_a(p^2)=\displaystyle\frac{-f^2}{(16\pi^2)}\int^1_0 d\alpha 
\, \ln \biggl[ \frac{\alpha^2
m^2+2(1-\alpha)\mu^2}{-\alpha(1-\alpha)p^2+\alpha m^2+2(1-\alpha)\mu^2} \biggr].
\eqno{(C.5)}
$$
The subtraction is so chosen that
$$
A_a(m^2)=B_a(m^2)=0.
$$

Expression (C.3) is not yet the answer for $\sum^{(r)}_a$, the subtraction for
which is given by (2.14).  Following the prescription of (2.14), we get
$$
{\scriptstyle\sum^{(r)}_a}(p)= \pslash A_a(p^2)+
mB_a(p^2)-2m^2[A^\prime_a(m^2)+B^\prime_a(m^2)](\pslash-m).
\eqno{(C.6)}
$$
From (C.4) and (C.5), we get
$$
A^\prime_a(m^2)=\displaystyle\frac{-f^2}{(16\pi^2)}\int^1_0
d\alpha\frac{\alpha(1-\alpha)^2}{\alpha^2m^2+2(1-\alpha)\mu^2},
$$
and
$$
B^\prime_a(m^2)=\displaystyle\frac{-f^2}{(16\pi^2)} \int^1_0 d\alpha
\displaystyle \frac{\alpha(1-\alpha)}{\alpha^2
m^2+2(1-\alpha)\mu^2}.
$$
It is then straightforward to find that, as $p\rightarrow\infty$,
$$
{\scriptstyle\sum^{(r)}_a}(p) \approx \frac{\pslash
f^2}{16\pi^2} \biggl[\frac{1}{2} \ln(-p^2/m^2)+\int^1_0 d\alpha\alpha
\frac{(4-4\alpha+\alpha^2)m^2-(2-\alpha)\mu^2}{\alpha^2
m^2+2\mu^2(1-\alpha)}-1 \biggr].
\eqno{(C.7)}
$$
\begin{center}
\bf {(2) Diagram (11b) }
\end{center}
The renormalized amplitude corresponding to diagram (11b) is
$$
\displaystyle{\scriptstyle\sum^{(r)}_b}(p)=if^2 \int \displaystyle\frac{d^4q}{(2\pi)^4}
\displaystyle\frac{\pslash-\qslash-m}{(p-q)^2-m^2} \displaystyle\frac{1}{q^2}-\mbox{term of
subtraction.}
\eqno{(C.8)}
$$
Comparing (C.8) with (C.1), we find that
$$
{\scriptstyle\sum^{(r)}_b}(p)= \pslash A_a(p^2)-m B_a(p^2)
-2m^2[A^\prime_a(m^2)-B_a(m^2)](\pslash-m), 
\eqno{(C.9)}
$$
with $\mu$ in $A_a$ and $B_a$ set to zero.  Thus we get, for
$p\rightarrow\infty$,
$$
{\scriptstyle\sum^{(r)}_b}(p) \equiv \displaystyle\frac{\pslash f^2}{(16\pi^2)}
\biggl[\frac{1}{2}
\ln(-p^2/m^2)-\frac{1}{2}\biggr]. 
\eqno{(C.10)}
$$

\begin{center}
\bf {(3)  Diagram (11c) }
\end{center}

Finally, we turn to diagram (11c).  The renormalized amplitude in the Landau
gauge corresponding to this diagram is
$$\begin{array}{ll}
{\scriptstyle\sum^{(r)}_c}(p) =& \displaystyle{-ig^2 \int \frac{d^4q}{(2\pi)^4}}
\displaystyle\frac{(3\qslash-2\pslash-\qslash\pslash\qslash/q^2) \displaystyle \frac{ 1+
2\theta+2\theta^2 + (1+2\theta)\gamma_5}{2} + 3m\theta(1+\theta)}{[(p-q)^2-m^2]
(q^2-M^2)} \\
 & - \mbox{term of subtraction.}
\end{array}
\eqno{(C.11)}
$$
Introducing Feynman parameters, making a shift of the variable, and carrying
out the integration, we get 
$$\begin{array}{ll}
{\scriptstyle\sum^{(r)}_c}(p)=& \displaystyle -g^2 \biggl[ A_c(p^2)\pslash 
\biggl(\frac{1+2\theta+2\theta^2 + (1+2\theta)\gamma_5}{2}\biggr)
+ m B_c(p^2) \theta (1+\theta) \biggr] \\
& -\mbox{term of
subtraction,} 
\end{array}
\eqno{(C.12)}
$$
where
$$\begin{array}{ll}
A_c(p^2)&\equiv \displaystyle{i\int^1_0 d\alpha\int}\displaystyle
\frac{d^4\bar{q}}{(2\pi)^4}\frac{(3\alpha-2)}
{[\bar{q}^2+\alpha(1-\alpha)p^2-\alpha
m^2-(1-\alpha)M^2]^2} \\
&\displaystyle{+i\int d^3\alpha \delta(1-\sum\alpha)\int}
\displaystyle\frac{d^4\bar{q}}{(2\pi)^4}\frac{\bar{q}-2\alpha^2_1p^2}{[\bar{q}^2+\alpha_1(1-\alpha_1)p^2-\alpha_1m^2-\alpha_2M^2]^3} .
\end{array}
\eqno{(C.13)}
$$
While each of the two integrals in (C.13) is logarithmically divergent, the
sum of the two integrals is ultraviolet finite.  Carrying out the integration
over $q$ we get, after some algebra,
$$
A_c(p^2)=\displaystyle\frac{1}{16\pi^2} \int^1_0
d\alpha \biggl[(2\alpha-1-\alpha\frac{p^2-m^2}{M^2})
\ln[-\alpha(1-\alpha)p^2+\alpha
m^2+(1-\alpha)M^2]
$$
$$+ \alpha \frac{p^2-m^2}{M^2} \ln[-\alpha(1-\alpha)p^2+\alpha m^2]-\alpha\biggr].
\eqno{(C.14)}
$$
From (C.14), it is straightforward to obtain
$$
A^\prime_c(m^2) = \frac{1}{16\pi^2} \int^1_0 d\alpha
\frac{\alpha^2(\displaystyle{-\frac{5}{2}}+2\alpha)}
{\alpha^2 m^2+(1-\alpha)M^2} .
\eqno{(C.15)}
$$
In a similar way, $B_c(p^2)$ and $B_c^\prime(m^2)$ are given by
$$
\begin{array}{ll}
B_c(p^2) & \displaystyle = i \int_0^1 d \alpha \int \frac{d^4 \bar{q}} {(2\pi)^4} \frac{3} {[\bar{q^2} + \alpha (1-\alpha) p^2 - \alpha m^2 - (1-\alpha) M^2]^2} - \mbox{term of subtraction} \\
& = \displaystyle \frac{3}{16\pi^2} \int_0^1 d \alpha \ln \biggl[ \frac{-\alpha(1-\alpha) p^2 + \alpha m^2 + (1-\alpha) M^2}{\alpha^2 m^2 + (1-\alpha) M^2} \biggr] , 
\end{array}
\eqno(C.16)
$$
$$
\displaystyle
B_c^\prime (m^2) = - \frac{3}{16\pi^2} \int_0^1 d \alpha \frac{\alpha(1-\alpha)}{\alpha^2 m^2 + (1-\alpha) M^2} .
\eqno(C.17)
$$
From (C.14), (C.15), (C.16) and (C.17), we get, following the prescription of (2.14),

$$\begin{array}{ll}
\displaystyle{\scriptstyle\sum^{(r)}_c}(p) &= -\displaystyle\frac{g^2}{16\pi^2}
\biggl[\displaystyle\int^1_0 d\alpha (2\alpha-1) \ln\displaystyle\frac{-\alpha(1-\alpha)p^2+\alpha m^2+(1-\alpha)M^2}{\alpha^2 m^2 +
(1-\alpha)M^2} \\
&+\displaystyle\alpha\displaystyle \frac{(p^2-m^2)}{M^2} \ln \displaystyle\frac{-\alpha(1-\alpha)p^2+\alpha
m^2}{-\alpha(1-\alpha)p^2+\alpha m^2+(1-\alpha)M^2} \biggr] \pslash
\displaystyle\frac{1+2\theta+2\theta^2 + (1+2\theta)\gamma_5}{2} \\
&+\displaystyle\frac{g^2 m^2}{16\pi^2} \int^1_0 d\alpha
\displaystyle\frac{\alpha^2(\displaystyle{-\frac{5}{2}}+2\alpha) (1+2\theta+2\theta^2) + 6\alpha(1-\alpha) \theta (1+\theta)}
{\alpha^2 m^2 +(1-\alpha)M^2} \, (\pslash - m)\\
& +\displaystyle \frac{3g^2 m}{16\pi^2} \theta(1+\theta)\int^1_0 d\alpha \, \ln \biggl[
\frac{\alpha^2 m^2 + (1-\alpha) M^2}{-\alpha(1-\alpha) p^2 + \alpha m^2 + (1-\alpha) M^2} \biggr] .
\end{array}
\eqno{(C.18)}
$$

The part of $\sum^{(a)}_c$ relevant for the anomaly amplitude is obtained from
(C.18) by taking the limit $p\rightarrow\infty$.
We find, after some
algebra, that the part in $\sum^{(r)}_c$ which contributes to the anomaly
amplitude is
$$\begin{array}{ll}
\displaystyle \frac {-g^2 \pslash} {32\pi^2} \int^1_0 d\alpha \frac{1}{\alpha^2 m^2 + (1 -\alpha) M^2} & \biggl\{
[2\alpha^2 (2-\alpha) m^2 + (1-\alpha^2) M^2](1+2\theta+2\theta^2) \\
& + [\alpha^2
(2\alpha-1) m^2 + (1-\alpha^2) M^2](1+2\theta) \gamma_5 \biggr\} \, .
\end{array} 
\eqno{(C.19)}
$$

\newpage
\begin{center}
\large \bf {Appendix D}
\end{center}

In this appendix we evaluate the ratio $Z_\psi (m) \bigr / Z_{V\psi\bar{\psi}}
(\Omega^2, \Omega^2, 0)$.  This quantity will be evaluated by the use of
(2.27).

  The values of $c_r(\Omega^2), d_r(\Omega^2)$ and $a_r(\Omega^2)$ have been
given in Appendices B and C, therefore we only need to calculate $G_+ (
\Omega^2, \Omega^2, 0)$.  After some calculations, we find that 
$$\begin{array}{ll}
\displaystyle
G_+ (\Omega^2, \Omega^2, 0) &= 
\displaystyle \frac{m f^2}{16\pi^2}  \int d^3 \alpha \delta (1-\sum\alpha) 
\displaystyle \biggl{\{}
\frac{1}{(\alpha_1+\alpha_2)m^2 - (\alpha_1+\alpha_2)\alpha_3\Omega^2
+ 2\alpha_3 \mu^2} \\
& {\hskip 0.5 cm}\displaystyle + \frac{1}{(\alpha_1+\alpha_2) m^2 - (\alpha_1+\alpha_2)^2
\Omega^2} + \frac{2\mu^2}{m^2} \frac{\alpha_2 - \alpha_1}{\alpha_3 m^2
+ 2\alpha_2\mu^2 - (\alpha_1+\alpha_2)\alpha_3\Omega^2} \biggr{\}} \\
\displaystyle 
& \displaystyle - \frac{m g^2}{16\pi^2}  \int d^3\alpha \delta (1-\sum\alpha) \biggl{\{}
\frac{2}{\alpha_1 m^2 - \alpha_1 (1-\alpha_1)\Omega^2 + 2\alpha_2\mu^2
+\alpha_3 M^2} \\
& {\hskip 3.5 cm}\displaystyle + \frac{1}{2} \frac{1}{(\alpha_1+\alpha_2)m^2 + 
\alpha_3 M^2 - (\alpha_1
+\alpha_2) \alpha_3 \Omega^2} \biggr{\}} \\
 \displaystyle
& \displaystyle - \frac{m g^2} {16\pi^2}  \int d^4\alpha \delta (1-\sum\alpha) \biggl{\{}
\frac{2}{\alpha_1 m^2 - \alpha_1 (1-\alpha_1)\Omega^2 + 2\alpha_2\mu^2
+\alpha_3 M^2} \\
& {\hskip 3.5 cm}\displaystyle + \frac{1}{2} \frac{1}{(\alpha_1+\alpha_2)m^2 + 
\alpha_3 M^2 - (\alpha_1
+\alpha_2) \alpha_3 \Omega^2} \biggr{\}}
\end{array}
\eqno(D.1)
$$

\newpage
\begin{center}
\large \bf {Appendix E}
\end{center}

In this section, we shall calculate the anomaly amplitude for $J_1$ and $J_2$.
The calculation of the volume integral $J_1$ will be presented in Section
(a), while the calculation of the surface integral $J_2$ will be 
presented in Section (b).

\noindent
{\bf (a) The Volume Integral $J_1$}

The Feynman integral corresponding to each of the diagrams in \fig7 is
quadratically divergent by power counting.  However, the trace in the
integrand of each integral vanishes if we set all of the momenta of
the external vector mesons to zero.  This means that each term in the
trace contains at least one factor of the external momenta and hence
the trace blows up no faster than the cubic power of the internal
momenta.  Consequently, each of the Feynman integrals in question is
linearly divergent.  As we have discussed in Section 3B as well
as in Section 5, the sum of these integrals are convergent by the requirement of Bose symmetry.

We shall first carry out the integration over $q$.  We shall do this
for the sum of anomaly amplitudes corresponding to diagrams (7a) and
(7b).  As we can conclude from power counting, this subintegration
over $q$ is convergent.  Next we carry out the integration over $p$.
As we have discussed in detail in Section 5, this integration is
also convergent as we sum over the contributions from all the diagrams.  The anomaly
amplitudes corresponding to diagrams (7c) and (7d) are equal to those
corresponding to diagrams (7a) and (7b) with the interchange of
$2\leftrightarrow3$.  We shall show that this sum is equal to
$$ 
J_1 = - \frac{1}{(8\pi^2)^2} g^2 \epsilon^{\nu \rho \sigma \sigma'}
k_{2\sigma} k_{3\sigma'} [ (1+\theta)^5 - \theta^5 ] ,
\eqno{(E.1)}
$$
a constant which depends on the renormalized coupling constant $g$, and
not on the masses of the particles in the theory.  This is because the
contributions to these anomaly amplitudes come from a region in which
the internal momenta are infinitely large, hence the particle masses
which appear in the integrand can be dropped.
We shall next present the calculation of these amplitudes.  Readers who
are not interested in this calculation are advised to skip the rest of
the subsection.

Referring to diagrams (7a) and (7b), we find the sum of amplitudes from
these two diagrams as
$$
M_A - M_B
=
i g^2
\int
  \frac{d^4 p}{(2\pi)^4}
  \frac{d^4 q}{(2\pi)^4}
\left(
  \frac{N_A}{\cal D_A} - \frac{N_B}{\cal D_B}
\right)
\eqno{(E.2)}
$$
where
$$\begin{array}{ll}
N_A
\equiv
Tr
\biggl[& \displaystyle
  \gamma^{\nu}
  \frac{1+2\theta+\gamma_5}{2}
  (\pslash + \kslash_2)
  \gamma^\sigma
  \frac{1+2\theta+\gamma_5}{ 2}
  (\pslash + \kslash_2 - \qslash)
  \gamma^\rho
  \frac{1+2\theta+\gamma_5}{2}
  (\pslash - \kslash_1 - \qslash) \\
  & \displaystyle \gamma^{\sigma'}
  \biggl(\frac{1+2\theta+\gamma_5}{2}\biggr)^2\pslash
\biggr]
\left(\displaystyle
  g_{\sigma\sigma'}
  -
  \frac{q_\sigma q_{\sigma'}}{q^2}
\right),
\end{array}
\eqno{(E.3)}
$$
\nopagebreak[4]
$$
{\cal D}_A
\equiv
  \left( p^2 - m^2 \right)
  \left[ (p + k_2)^2 - m^2 \right]
  \left[ (p + k_2 - q)^2 - m^2 \right]
  \left[ (p - k_1 - q)^2 - m^2 \right]
  \left( q^2 - M^2 \right)
,
\eqno{(E.4)}
$$ 
  and similarly for $N_B$ and $\cal D_B$.  We have dropped the
  fermion mass $m$ in the numerators of the fermion propagators.  This
  is because the terms that involve $m$ in the numerator are
  convergent integrals.  Thus, shifting the loop-momenta for the terms
  which have $m$ in the numerator is allowed and hence such terms in $M_A$
  cancel the corresponding terms in $M_B$.  The propagator of the vector meson is in the
  Landau gauge, which accounts for the last factor in (E.3).  As
  noted, the contribution of the left-handed fermion and the
  right-handed fermion can be calculated separately, as all five
  vertices on the fermion loop are $V\psi\bar{\psi}$ vertices.  Thus
$$ 
N_A
\equiv [ (1+\theta)^5 - \theta^5 ] 
\biggl( N_A^{(1)}
+
\half N_A^{(2)}
/
q^2 \biggr)
,
\eqno{(E.5)}
$$
with
$$
N_A^{(1)}
=
Tr
\biggl[
  \gamma_5
  \gamma^{\nu}
  (\pslash + \kslash_{2})
  (\pslash - \kslash_{1} - \qslash)
  \gamma^\rho
  (\pslash + \kslash_{2} - \qslash)
  \pslash
\biggr]
,
\eqno{(E.6a)}
$$
and
$$
N_A^{(2)}
=
\biggl[
  \gamma_5
  \gamma^{\nu}
  (\pslash + \kslash_{2})
  \qslash
  (\pslash + \kslash_{2} - \qslash)
  \gamma^\rho
  (\pslash - \kslash_{1} - \qslash)
  \qslash
  \pslash
\biggr]
,
\eqno{(E.6b)}
$$
where we have averaged over the two directions of the fermion loop.  We
shall put
$$
M_A
\equiv
M_A^{(1)}
+
M_A^{(2)}
,
$$
where $M_A^{(1)}\biggl ( M_A^{(2)} \biggr)$ is obtained from $M_A$ by replacing
 $N_A$ with 
with $N^{(1)}_A \biggl( \displaystyle \frac{1}{2} N_A^{(2)}/q^2 \biggr) $. 
We also define $M_B^{(1)}$ and
$M^{(2)}_B$ similarly with reference to diagram (7b).  By introducing Feynman parameters and carrying out the
integration over $q$, we get, after some algebra,
$$
\vbox{\ialign{%
$\displaystyle\hfil#$&
$\displaystyle\hfil\null#\null\hfil$&
$\displaystyle#\hfil$\cr
  M^{(1)}_A-M^{(1)}_B&
  =&
  -(8\pi^2)^{-1}g^2 \int d^5\alpha
  \delta(1-\sum\alpha)\int\frac{d^4p}{(2\pi)^4} \frac{1}{\wedge^3}\cr
  &&
  \left\{
  \frac{rN_A^{(1)}}%
  {\displaystyle
    \left[
      (p-\Delta_A)^2 + r D_A \wedge^{-2}
    \right]^3}
  -
  \frac{\alpha_3\alpha_5
    Tr(\gamma_5 \gamma^\nu \gamma^\rho \kslash_2\kslash_3)}%
  {\displaystyle 2r
    \left[
      (p-\Delta_A)^2 + r D_A \wedge^{-2}
     \right]^2}\right.\cr
\noalign{\smallskip}
  &&\quad\null
  -
  \left.
    \frac{rN_B^{(1)}}{\displaystyle
      \left[
        (p-\Delta_B)^2 + r D_B \wedge^{-2}
      \right]^3}
    -
    \frac{\alpha_4 \alpha_5
      Tr(\gamma_5\gamma^\nu\gamma^\rho \kslash_2\kslash_3)}%
    {\displaystyle
      2r \left[
          (p-\Delta_B)^2 + r D_B \wedge^{-2}
        \right]^2}
  \right\}.\cr}}
\eqno{(E.7)}
$$
In (E.7),
$$
\begin{array}{lll}
  D_A &
  \equiv &
  \alpha_1\alpha_3\alpha_5 k^2_1\alpha_1 +
  \left[
    \alpha_5(\alpha_2+\alpha_4) + \alpha_2(\alpha_3+\alpha_4)
  \right]
  k^2_2+\alpha_3
  \left[
    (\alpha_1+\alpha_2)\alpha_4+(\alpha_2+\alpha_4)\alpha_5
  \right]
  k^2_3\\
  &&\null
    + \left [(\alpha_1+\alpha_2+\alpha_3+\alpha_4)m^2+\alpha_5 M^2
  \right]\wedge,\\
  \wedge &
  \equiv &
  (\alpha_1+\alpha_2) (\alpha_3+\alpha_4)
  +
  \alpha_5(\alpha_1+\alpha_2+\alpha_3+\alpha_4),\\
  r &
  \equiv &
  \alpha_3+\alpha_4+\alpha_5;
\end{array}$$
$$\eqno{(E.8)}$$
also,
$$
N^{(1)}_A
\equiv
Tr\left[
\gamma_5\gamma^\nu(\pslash_A+\pslash_1)
(\frac{\alpha_5}{r}\pslash_A+\pslash_2)
\gamma^\rho
(\frac{\alpha_5}{r}\pslash_A+\pslash_3)
(\pslash_A+\Deltaslash_A)
\right],
\eqno{(E.9)}
$$
with
$$
p_A \equiv p-\Delta_A,
\eqno{(E.10)}
$$
and
$$
\begin{array}{lll}
\wedge p_1 & \equiv &
\alpha_1(\alpha_3+\alpha_4+\alpha_5)k_2-\alpha_3\alpha_5k_3,\\
\wedge p_2 & \equiv & \alpha_1\alpha_5 k_2
+[(\alpha_1+\alpha_2)(\alpha_4+\alpha_5)+\alpha_4\alpha_5]k_3,\\
\wedge p_3 & \equiv & \alpha_1\alpha_5 k_2 -
\alpha_3(\alpha_1+\alpha_2+\alpha_5)k_3,\\
\wedge \Delta_A & \equiv &
-[(\alpha_2+\alpha_5)(\alpha_3+\alpha_4)+\alpha_2\alpha_5]k_2-\alpha_3\alpha_5k_3.
\end{array}
$$

In addition, $D_B$ is equal to $D_A$ with $\alpha_1\leftrightarrow\alpha_2$,
$\alpha_3\leftrightarrow\alpha_4$:
$$
D_B = D_A\bigg|_{%
\displaystyle\alpha_1\leftrightarrow\alpha_2,\alpha_3\leftrightarrow\alpha_4},
$$
\nopagebreak[4]
$$
N^{(1)}_B =
Tr\left[
\gamma_5\gamma^\nu(\pslash_B+\pslash^\prime_1)
(\frac{\alpha_5}{r}\pslash_B+\pslash_2^\prime)
\gamma^\rho(\frac{\alpha_5}{r}\pslash_B+\pslash^\prime_3)
(\pslash_B+\Deltaslash_B)
\right],
\eqno(E.11)
$$
with
$$
\begin{array}{lll}
p_B & \equiv  & p-\Delta_B,\\
\wedge p^\prime_1 & \equiv & \alpha_4\alpha_5 k_3
+[(\alpha_1+\alpha_5)(\alpha_3+\alpha_4)+\alpha_1\alpha_5]k_2,\\
\wedge p^\prime_2 & \equiv &
-\alpha_2\alpha_5k_2+\alpha_4(\alpha_1+\alpha_2+\alpha_5)k_3,\\
\wedge p^\prime_3 & \equiv &
-\alpha_2\alpha_5k_2+[\alpha_3(\alpha_1+\alpha_2)+\alpha_5(\alpha_1+\alpha_2+\alpha_3)]k_3,\\
\wedge\Delta_B & \equiv & \alpha_4\alpha_5 k_3 -
\alpha_2(\alpha_3+\alpha_4+\alpha_5)k_2.
\end{array}
$$
$$\eqno(E.12)$$

Next we carry out the integration over $p$ in (E.7).  Consider the first
term of the integrand in (E.7).  We shall change the variable of
integration from $p$ to $p_A$ which is defined in (E.10).  This is to
say that, instead of performing a symmetric integration over $p$, we shall
perform a symmetric integration over $p_A$.  Since $N^{(1)}_A$ given by
(E.9) is cubic in $p_A$, the corresponding integral is linearly
divergent, and a term proportion to $\Delta_A$ is generated by the
change of variables.  In addition, we note that the term in $N^{(1)}_A$
obtained by setting $p_A$ to zero is exactly canceled by its counterpart
in $M^{(1)}_D$.  This is because diagram (7a) is topologically identical
to diagram (7d), and the amplitudes corresponding to these two diagrams
are different only because the loop-momenta are chosen differently.  But
$p_1$ in $N^{(1)}_A$ is equal to its counterpart in $N^{(1)}_D$.  To see
this, let us think of diagram (7a) as an electrical circuit, with
$\alpha_i$ the resistance of line $i$.  Then $p_1$ for example is the
current in line 2.  Since diagram (7d) represents the same electrical
circuit, it has the same currents in its lines.  This is why $N^{(1)}_A$
at $p_A=0$ is exactly equal to $N^{(1)}_D$ at $p_D=0$.  Thus we shall
ignore such a term in $N^{(1)}_A$.  Similarly, we shall ignore the term
in $N^{(1)}_B$ which is independent of $p_B$.  After dropping such terms
and calling both $p_A$ and $p_B$ as
$\bar{p}$, the bracket in (E.7) becomes, after some algebra and changes of
variables, 
$$
\frac{\bar{p}^2\alpha_5[2\alpha_3(\alpha_2+\alpha_5)-2\alpha_1(\alpha_4+\alpha_5)-\alpha_1
r+\alpha_3\alpha^2_5/r]}{\wedge(\bar{p}^2+D_a r/\wedge^2)^3}
+\frac{\alpha_3\alpha_5}{r(\bar{p}^2+D_a r/\wedge^2)^2}.
\eqno(E.13)
$$
Substituting (E.13) into (E.7) and carrying out the integration over
 $\bar{p}$, and adding to it $M^{(1)}_C-M_D^{(1)}$ which is obtained from
 $M^{(1)}_A - M_B^{(1)}$ by the interchange of $2\leftrightarrow3$, we
get
$$
\begin{morelines}
  M^{(1)}_A-M^{(1)}_B +M^{(1)}_C-M^{(1)}_D \hidewidth\\
\qquad  &=&
  2ig^2(16\pi^2)^{-2}
  Tr(\gamma_5\gamma^\nu\gamma^\rho \kslash_2 \kslash_3)
  \int
  \frac{d^5\alpha\delta(1-\sum\alpha)}{\wedge^4}\\
  &&
  \left[
    \alpha_1\alpha_5(\alpha_3+\alpha_4+\alpha_5)
    -
    \alpha_3\alpha^3_5(\alpha_3+\alpha_4+\alpha_5)
    +
    3\alpha^2_5(\alpha_3-\alpha_1)
    ln
    \left(
      \frac{\alpha_1+\alpha_2+\alpha_5}{\alpha_3+\alpha_4+\alpha_5}
    \right)
  \right.\\
\noalign{\smallskip}
  &&
  ~~\null
  -
  \left.
  \frac{\alpha_5^2(\alpha_3+\alpha_4)(\alpha_3+\alpha_4+2\alpha_5)}%
       {\alpha_3+\alpha_4+\alpha_5}
  -
  \frac{1}{2}
  \frac{\wedge\alpha_5(\alpha_3+\alpha_4)}{\alpha_3+\alpha_4+\alpha_5}
  \right],
\end{morelines}
\eqno{(E.14)}
$$
with the last two terms in the integrand derived from the shifting of the
integration variables.  The integrations over the Feynman parameters are
routine and we find after some algebra that the right-side of (E.14) is equal to the expression
in (E.1).

Next we calculate $M_A^{(2)} - M_B^{(2)}$.
We get
$$
M^{(2)}_A-M^{(2)}_B = \frac{1}{2}ig^2\int
\frac{d^4p}{(2\pi)^4}\frac{d^4q}{(2\pi)^4}\frac{1}{q^2}
\left[
  \frac{N^{(2)}_A}{\cal D_A} -
  \frac{N^{(2)}_B}{\cal D_B}
\right],
$$
with $N^{(2)}_A$ given by (E.6b), and with $N^{(2)}_B$ given by a similar
expression with reference to diagram (7b). 
 
The ensuing calculations are somewhat simplified if we replace the first
$\qslash$ in (E.6b) by
$$
(\pslash+\kslash_2)-(\pslash+\kslash_2-q) .
$$
Then we get
$$
\begin{morelines}
  N^{(2)}_A&
  =&
  (p+k_2)^2
  Tr
  \biggl[
    \gamma_5\gamma^\nu(\pslash+\kslash_2-\qslash)
    \gamma^\rho       (\pslash-\kslash_1-\qslash)
    \qslash\pslash
  \biggr]\\
  &&\quad\null
  -
  (p+k_2 -q)^2Tr
  \biggl[
    \gamma^\nu (\pslash+\kslash_2)
    \gamma^\rho(\pslash-\kslash_1-\qslash)
    \qslash\pslash
  \biggr] .
\end{morelines}
\eqno{(E.15)}
$$
Now
$$
(p+k_2)^2=[(p+k_2)^2-m^2]+m^2.
\eqno(E.16)
$$
The first term on the right-side of (E.16) will be used to cancel the same
factor in $\cal D_A$.  The second term in (E.16), $m^2$, will be dropped.  This
term is by two powers of loop-momenta less divergent than the first term, and
its contributions are cancelled by its counterpart of diagram (7d).  Next we
introduce Feynman parameters, integrating first over $q$ and then symmetrically over $p$.  In the latter
integration, a change of variable from $p$ to $\bar{p}$ is made.  The part
of $\displaystyle M^{(2)}_A-M^{(2)}_B+M^{(2)}_C-M^{(2)}_D$ which comes from
symmetric integration over $\bar{p}$ is calculated to be
$$
\begin{morelines}
  &\hss&
  \frac{2ig}{(16\pi^2)^2}
  Tr(\gamma_5 \gamma^\nu \gamma^\rho \kslash_2 \kslash_3)
  \int
  \frac
  {d\alpha_1 d\alpha_2 d\alpha_3 d\alpha_5 \delta(1-\sum\alpha)\alpha^2_5}%
  {\left[
      (\alpha_1+\alpha_2)\alpha_3
      +\alpha_5(\alpha_1+\alpha_2+\alpha_3)
   \right]^4}\\
  \noalign{\smallskip}
  &\hss&
  \qquad
  \left[
    -3\alpha_3(\alpha_1+\alpha_2)
    \ln
    \left(
      \frac{\alpha_1+\alpha_2+\alpha_5}{\alpha_3+\alpha_5}
    \right)
    +
    \frac{\alpha_1(\alpha_1+\alpha_2)\alpha_5}{\alpha_1+\alpha_2+\alpha_5}
    -
    \frac{\alpha^2_3\alpha_5}{\alpha_3+\alpha_5}
    -
    \alpha_2\alpha_3
  \right].
\end{morelines}
\eqno{(E.17)}
$$
It is straightforward to carry out the integration over the Feynman
parameters.  We find that the expression in (E.17) is equal to
$$
-\frac{1}{2}ig^2(16\pi^2)^{-2}Tr(\gamma_5\gamma^\nu\gamma^\rho \kslash_2
\kslash_3).
\eqno(E.18)
$$
In addition, there is a term which comes from the change of integration
variables from $p$ to $\bar{p}$.  This term is equal
to the negative of (E.18).  Thus
$$
M^{(2)}_A-M^{(2)}_B+M^{(2)}_C-M^{(2)}_D=0.
\eqno(E.19)
$$
Consequently, the sum of anomaly amplitudes corresponding to the diagrams in
Fig. 7 is
$$
  [ (1+\theta)^5 - \theta^5 ] [ M_A - M_B + M_C - M_D ]
  =
 - \frac{1}{(8\pi^2)^2}g^2
  \epsilon^{\displaystyle\nu\rho\sigma\sigma'}\displaystyle
  k_{2\sigma} k_{3\sigma'} [ (1+\theta)^5 - \theta^5 ] .
\eqno(E.20)
$$

\noindent
{\bf (b) The Surface Term $J_2$}

        In this subsection we evaluate the surface integral $J_2$ given by (4.5)
in the two lowest orders of the renormalized coupling constants.

First we express the unrenormalized functions $\Gamma_{V\psi\bar{\psi}}$ and
$S$ in (4.5b) by the corresponding renormalized ones.  Then $N_7$ is given
by (4.5b) with $\Gamma$ and $S$ replaced by $\Gamma^{(r)}$ and $S^{(r)}$,
respectively, with an additional overall factor
$$
\biggl[ Z_\psi (m) \biggr/ Z_{V\psi\bar{\psi}} (\Omega^2, \Omega^2, 0)
\biggr]^2 .
\eqno(E.21)
$$
All of the fermion momenta at the surface of integration of (4.5a) are 
infinitely large.  Therefore, we only need to consider the asymptotic forms
of $\Gamma$ and $S$ in this limit.  Referring to 
(2.2) and (2.12), we have, for $p^2 
\rightarrow \infty$,
$$
 L {[S^{(r)} (p) ]}^{-1}  \approx  -i L 
\pslash (1 - c_r + d_r)  ,
\eqno(E.22a)
$$
$$ 
 {[ S^{(r)} (p)]}^{-1} L \approx -i \pslash (1-c_r-d_r) L , 
\eqno(E.22b)
$$
and
$$
 R {[S^{(r)} (p)]}^{-1} \approx -i R \pslash(1-c_r-d_r) ,  
\eqno(E.23a)
$$
$$
 {[S^{(r)} (p)]}^{-1} \approx -i \pslash(1 - c_r + d_r) R   .
\eqno(E.23b)
$$
where $L$ and $R$ are defined in (3.16).
As to $\Gamma_{V\psi\bar{\psi}}^{\nu}$, it is a linear superposition of
a number of invariant amplitudes each multiplied by one of the following
terms:
$$
\displaystyle {\gamma^{\nu} \frac{1\pm\gamma_5}{2} \, ,\, m \Pslash \gamma^{\nu}
\frac{1\pm\gamma_5}{2} \, , \,m \Deltaslash \gamma^{\nu} \frac{1\pm\gamma_5}{2}
\, , \,\pslash'\gamma^{\nu} \pslash \frac{1\pm \gamma_5}{2}},
$$
$$ 
\displaystyle {m\Delta^{\nu} \frac{1\pm\gamma_5}{2}\, ,\, \Pslash\Delta^{\nu}
\frac{1\pm\gamma_5}{2}\, ,\, \Deltaslash \Delta^{\nu} \frac{1\pm\gamma_5}{2} 
\, , \, m\pslash' \Delta^{\nu} \pslash \frac{1\pm\gamma_5}{2} },
$$
$$ 
\displaystyle { mP^{\nu} \frac{1\pm\gamma_5}{2}\, ,\, \Pslash P^{\nu} \frac
{1\pm\gamma_5}{2}\, ,\, \Deltaslash P^{\nu} \frac{1\pm \gamma_5}{2} \,
, \, m\pslash'
P^{\nu} \pslash \frac{1\pm\gamma_5}{2} } .
$$
$$
\eqno(E.24)
$$
All terms in (E.24) are of the dimension of even powers of momentum.
In this way, the invariant amplitudes associated with these terms are also
of the dimension of even powers of momentum, and are functions of $p^2, 
p^{'2}$ and $\Delta^2$.

The function $\Gamma_{V\psi\bar{\psi}}$ is dimensionless.  Therefore, in the 
limit of $p^2$ and $p^{'2}$ going to infinity with $\Delta^2$ fixed, the
surviving terms in (E.24) are
$$
\displaystyle{\gamma^{\nu}\frac{1\pm\gamma_5}{2} \, , \, \pslash' \gamma^{\nu}
\pslash \frac{1\pm\gamma_5}{2}\, ,\, \Pslash P^{\nu} \frac{1\pm\gamma_5}{2} }.
\eqno(E.25)
$$
For example, the coefficient of $\displaystyle m\Pslash \gamma^{\nu} 
\frac{1\pm\gamma_5}{2}$
is by dimensional considerations of the order of $(P^2)^{-1}$ as $P^2
\rightarrow \infty$ .  Since $(P^2)^{-1} m \Pslash \gamma^{\nu} \displaystyle
\frac{1\pm\gamma_5}{2}$ vanishes as $P^2 \rightarrow \infty$,
this amplitude can be neglected.  Thus, all the terms in (E.24) with a factor
$m$ can be neglected.  Similarly, all the terms in (E.24) with a factor 
$\Delta$ can be neglected.   
We may also show that the term $\Pslash P^{\nu}\displaystyle
\frac{1+\gamma_5}{2}$ in (E.25) does not 
contribute to $J_2$, as it is possible to verify that $N_7$ vanishes asymptotically if we substitute 
such a term for 
$\Gamma_{V\psi\bar{\psi}}$ in (4.5b).  Thus, the terms in $\Gamma_{V
\psi\bar{\psi}}^{(r)}$ contributing to $J_2$ are the first two terms of (E.25).
We shall therefore put $\Gamma_{V\psi\bar{\psi}}^{(r)}$ in the form
$$
\displaystyle {(1+\alpha_+) \gamma^{\nu} (1+\theta)\frac{1+\gamma_5}{2} 
+ (1+\alpha_-) \gamma^{\nu} \theta \frac{1-\gamma_5}{2} + \tau_+
\frac{\pslash' \gamma^{\nu} \pslash }{p^2} (1+\theta)\frac{1+\gamma_5}{2} 
+\tau_- \frac{\pslash' \gamma^{\nu} \pslash}{p^2} \theta \frac{1-\gamma_5}{2} } , 
\eqno(E.26)
$$
Substituting
(E.23) and (E.26) into (4.5b), we find after some algebra that, in the
two lowest orders in the coupling constant
$$\begin{array}{ll}
\displaystyle
J_2 = \epsilon^{\nu\rho\sigma\sigma'} \frac{k_{1\sigma}k_{2
\sigma'}}{12\pi^2} \biggl[ \frac{Z_{\psi} (m)}{Z_{V\psi\bar{\psi}} (
\Omega^2, \Omega^2, 0) } \biggr]^2 & \displaystyle 
\biggl\{ (1+\theta)^3 [(1 + 2\alpha_+ 
-  \frac{\tau_+}{2} + 2c_r + 2d_r ] \\
& \displaystyle - \theta^3 [ 1+ 2\alpha_- - \frac{\tau_-}{2} + 2c_r - 2d_r ] \biggr\} . 
\end{array}
\eqno(E.27)
$$

The asymptotoic forms for $\alpha_{\pm}, c_r$ and $d_r$ have been evaluated explicitly in Appendix 
B and C, while the value of $Z_{\psi} (m) \bigr / Z_{V\psi\bar{\psi}} 
(\Omega^2, \Omega^2, 0) $ was evaluated explicitly in Appendix D.  The calculations
in these appendices are straightforward but tedious.  Such explicit calculations
can be avoided if we make use of the Ward-Takahashi identity (2.19) in the
limit of $p^2$ and $p^{'2}$ going to infinity with $\Delta^2$ fixed.  Let
$$
\displaystyle \frac{Z_{\psi} (m)} {Z_{V\psi\bar{\psi}} (\Omega^2, \Omega^2, 0)}
\equiv 1 + z .
\eqno(E.28)
$$
Substituting (E.26) into the Ward-Takahashi identity (2.19) and making use of
(E.22) and (E.23), we can equate the 
coefficients of $\displaystyle \Deltaslash \frac{1 \pm \gamma_5}{2}$ and
obtain, in the
limit of $p^2$ and $p^{'2}$ going to infinity with $\Delta^2$ fixed,
$$
z + \alpha_{\pm} - \tau_{\pm} + c_r \pm d_r = 0 .
\eqno(E.29)
$$
Substituting (E.29) into (E.27), we get
$$
\displaystyle 
J_2 = \epsilon^{\nu\rho\sigma\sigma'} \frac{k_{1\sigma} k_{2\sigma'}} {12\pi^2}
\biggl[ (1+\theta)^3 - \theta^3 + \frac{3}{2} (1+\theta)^3\tau_+ - \frac{3}{2}
\theta^3 \tau_- \biggr] . 
\eqno(E.30)
$$
In Appendix B, we showed that the only diagram which yields an 
amplitude $\tau_+$ are diagrams (6a), (6b) and (6c).  The value of $\tau_{\pm}$ are given by 
 
$$\displaystyle
\tau_+ = (1+\theta)^2 \frac{g^2}{8\pi^2} - \frac{\theta}{1+\theta} \frac{f^2}
{8\pi^2} \,\, , 
\hskip 1 cm 
\tau_- = \theta^2 \frac{g^2}{8\pi^2} - \frac{1+\theta}{\theta} \frac{f^2}
{8\pi^2} \,\, . 
\eqno(E.31)
$$

\newpage
\begin{center}
\large \bf {Acknowledgements}
\end{center}

One of us (H.C.) would like to thank Professor Daniel Freedman and
Professor T.T. Wu for discussions.

\newpage
\begin{center}
\large \bf {References}
\end{center}
 
\begin{enumerate}

\item {H. Cheng and S.P. Li, Physical Masses and the Vacuum Expectation Value
of the Higgs field (to be published).}

\item {J. Ashmore, Lett. Nuovo Cimento \underline {4}, 289 (1972); C.G.
Bollini and J.J. Giambiagi, Phys. Lett \underline {40B}, 566 (1972).}

\item {G. 't Hooft and M. Veltman, Nucl. Phys. \underline{B44}, 189, (1972).}

\item {For a review of this issue, see, for example, Guy Bonneau,
International Journal of Modern Physics, Vol. 5, \underline{3831} (1990). See
also, A. Barroso et. al., Phys. Lett. B261, \underline{123} (1991).}

\item {See, for example, Sec. 6.4 in S.J. Chang, Introduction to Quantum Field
Theory (World Scientific, 1990) for a simple discussion on the BPHZ
renormalization.}

\item{H. Cheng and E.C. Tsai, Chinese J. of Phys., Vol. \underline{25}, No. 1,
95 (1987), Phys. Rev. Lett., \underline{57}, 511 (1986); Phys. Rev. \underline
{D40}, 1246 (1989).}

\item{We note that, since $(k_1+k_2+k_3)=0$, the amplitude $W_1$ is not
uniquely defined.  For example, we may add to $W_1(k^2_1,k^2_2,k^2_3)$ an
amount $(k^2_1-k^2_2)(k^2_2-k^2_3)(k^2_3-k^2_1)$, which is invariant under
cyclic permutation, without changing the value of the amplitude
$\Gamma^{\mu\nu\rho}_{VVV}$.  Nevertheless, (3.7) is valid for any choice of
$W_1$ extracted from an amplitude in the form (3.2), as long as it does not
have unwarranted singularities at $k^2_i=0, i=1,2,3$.  Similarly, $W_2, W_3$
are $W_4$ are not uniquely defined.  Among others, we have
$$\displaystyle\epsilon^{\mu\nu\rho\sigma}[(k_2\cdot k_2)k_{3\sigma}-(k_1\cdot k_3)k_{2\sigma}]=[k^\mu_1
\epsilon^{\nu\rho\sigma\sigma'}-k_1^\mu
\epsilon^{\mu\rho\sigma\sigma'}+k^\rho_1
\epsilon^{\mu\nu\sigma\sigma'}]k_{2\sigma}k_{3\sigma'},$$ 
which can be obtained from moving $\kslash_1$ in
$Tr(\kslash_1\kslash_2\kslash_3\gamma^5\gamma^\mu\gamma^\nu\gamma^\rho)$ to
the right, one step at a time, until it reaches the rightest postion.}

\item {S.L. Adler, Phys. Rev. \underline{177}, 2426 (1969).}

\item{J.S. Bell and R. Jackiw, N. Cimento, \underline{60A}, 47 (1969).}

\item {For an exposition on the anomaly, see, for example, Kerson Huang,
Quarks, Leptons and Gauge Fields (World Scientific)1992 and reference
therein.}

\item {In our notation, $\displaystyle\gamma_5 = -
i\gamma^0\gamma^1\gamma^2\gamma^3$, and $\displaystyle
Tr(\gamma_5A\!\!\!/B\!\!\!/C\!\!\!/D\!\!\!/) =
4i\epsilon^{\mu\nu\rho\sigma}A_\mu B_\nu C_\rho D_\sigma.$}

\item{S.L. Adler and W.A. Bardeen, Phys. Rev. \underline{182}, 1517 (1969).}
 
\item{This choice of notations was found by T.T. Wu (private communications).  An equivalent choice of notations is given by D.J. Gross and R. Jackiw, Phys. Rev. \underline{D6}, 477 (1972).}

\item{This result is in agreement with Joshua Erlich and Dan Freedman, Conformal
Symmetry and the Chiral Anomaly(preprint, 18 Nov., 1996) who calculated Abelian
as well as non-Abelian cases with zero masses.  This approach is based on 
calculatoins in the configuration space and is complementary to ours which is inthe momentum space.  H.C. appreciates his sharing their result with us.}

\end{enumerate} 

\newpage
\begin{center}
\large \bf {Figure Captions}
\end{center}
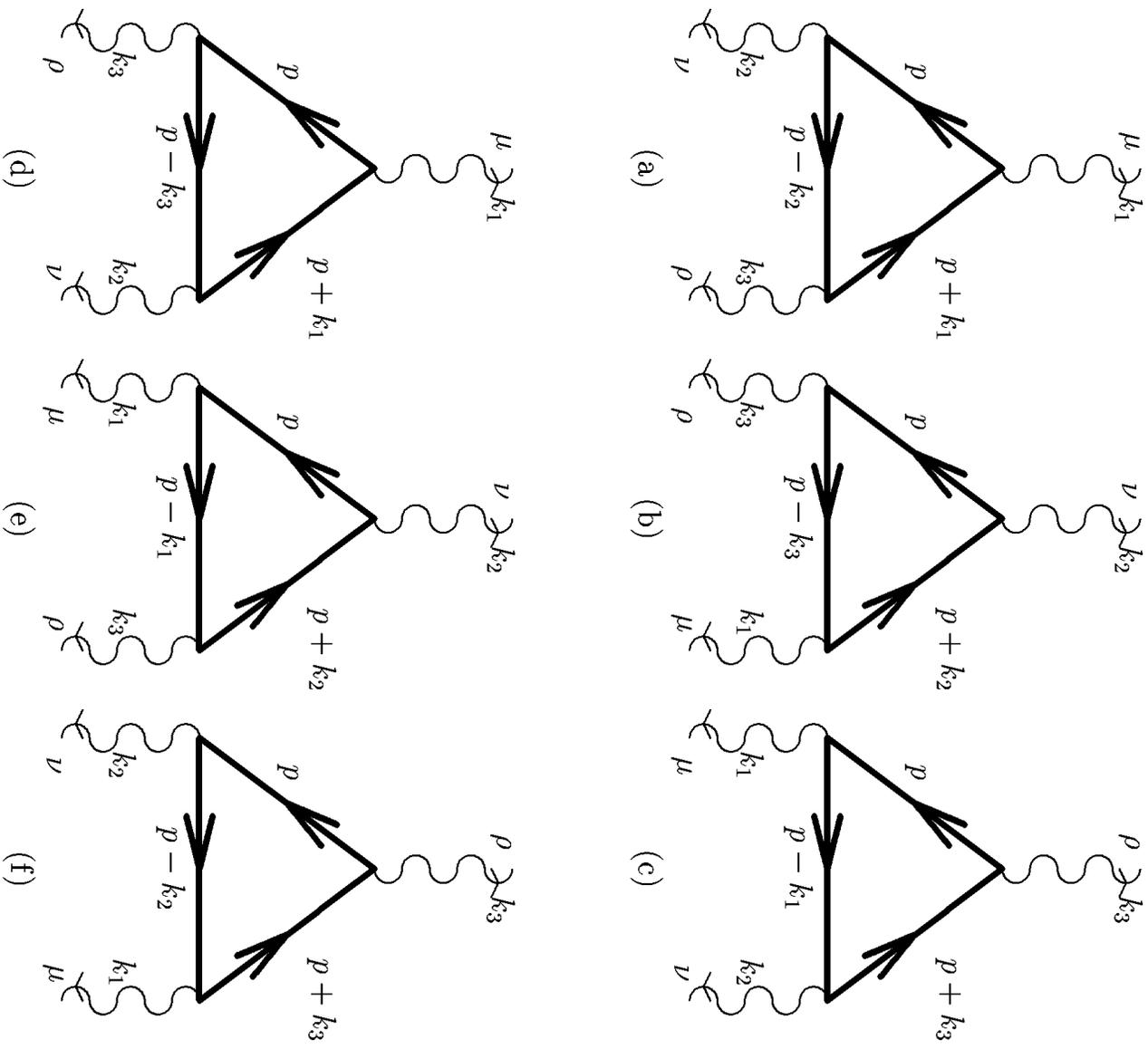
Figure 1.  Lowest-order triangular graphs.  The thick straight lines
represent fermions, and

\hskip 1.1 cm
the wavy lines represent vector
mesons.  Only two of the six diagrams are 

\hskip 1.1 cm
topologically independent.

\vskip 0.3 cm
\noindent
Figure 2.  A double line represents a $\phi_2$-meson.
See the caption of Figure 1 for the meanings

\hskip 1.1 cm
of other lines.

\vskip 0.3 cm
\noindent
Figure 3.  A vertex with a blob represents a dressed
vertex, with all of the radiative corrections

\hskip 1.1 cm
included.  A line with a blob represents a
dressed propagator.

\vskip 0.3 cm
\noindent
Figure 4.  A straight thin line represents a H-meson.
For meanings of other lines, see caption

\hskip 1.1 cm
of Figure 1.

\vskip 0.3 cm
\noindent
Figure 5.  A diagram in this figure represents an
anomaly amplitude.  A dotted line in 

\hskip 1.1 cm
these
diagrams represents the vector meson of
momentum $k_1$, the vertex for 

\hskip 1.1 cm
which is
absent as the Ward-Takahashi relation
(2.17) has been applied.

\vskip 0.3 cm
\noindent
Figure 6.  The diagrams for the lowest-order radiative
                   corrections of the $V\psi\bar{\psi}$ vertex.

\vskip 0.3 cm
\noindent
Figure 7.  The anomaly diagrams obtained from those in
Figure 5 with a dressed vertex 

\hskip 1.1 cm
replaced by
a vertex graph in Figure 6a.

\vskip 0.3 cm
\noindent
Figure 8.  The anomaly diagrams obtained from those in 
Figure 5 with a dressed vertex 

\hskip 1.1 cm
replaced by
a vertex graph of Figure 6b.

\vskip 0.3 cm
\noindent
Figure 9.  The set of nonplanar diagrams which contributes to the next order

\hskip 1.1 cm
triangle anomaly.

\vskip 0.3 cm
\noindent
Figure 10.  Diagrams depicting the reduction of amplitudes
by the use of (4.8).  A dot on  

\hskip 1.1 cm
an external
vector-meson line represents the multiplication
of $ik_\mu$ to the $V\psi\bar{\psi}$  

\hskip 1.1 cm
amplitude in (4.8).

\vskip 0.3 cm
\noindent
Figure 11. Same as Figure 10 but with mesons 2 and 3 interchanged.

\vskip 0.3 cm
\noindent
Figure 12. The anomaly diagrams obtained from those in Figure 5 with a 
dressed vertex

\hskip 1.1 cm
replaced by a vertex graph in Figures (6f) and (6g). 

\vskip 0.3 cm
\noindent
Figure 13. The anomaly diagrams obtained from those in Figure 9 which
contribute to

\hskip 1.1 cm
the next order triangle anomaly amplitudes.

\vskip 0.3 cm
\noindent
Figure 14. Diagrams representing three anomaly amplitudes
                   which cancel one another.

\vskip 0.3 cm
\noindent
Figure 15. Lowest-order radiative corrections for the
                   fermion propagator.

\end{document}